\documentclass[3p,authoryear]{elsarticle}

\usepackage[authoryear]{natbib}
\usepackage{amsfonts,amsmath,graphicx}
\usepackage{multirow}
\usepackage[usernames,dvipsnames]{xcolor}

\newcommand{\zOfT}{t}
\newcommand{\mur}[1]{\mathrm{#1}}

\newcommand{\mathd}{\mathrm{d}}
\newcommand{\mathpi}{\pi}
\newcommand{\nobracket}{}

\newcommand{\tmem}[1]{{\em #1\/}}
\newcommand{\tmmathbf}[1]{\ensuremath{\boldsymbol{#1}}}

\newcommand{\mymult}{\,}
\begin{document}	
	\journal{Submitted to R. Proc Soc A}

	\begin{frontmatter}	
	\title{A one-dimensional model for elasto-capillary necking}
	
	\author{Claire Lestringant$^{1}$ and Basile Audoly$^{2}$}
	
	\address{$^{1}$Structures Research Group, Department of Engineering, University of Cambridge, Cambridge CB2 1PZ, United Kingdom, casl4@cam.ac.uk\\
		$^{2}$Laboratoire de M\'ecanique des Solides, CNRS, Institut 
		Polytechnique de Paris, 91120 Palaiseau, France}
		
		\begin{abstract}
			We derive a non-linear one-dimensional (1d) strain
			gradient model predicting the necking of soft elastic
			cylinders driven by surface tension, starting from 3d
			finite-strain elasticity.  It is asymptotically correct:
			the microscopic displacement is identified by an energy
			method.  The 1d model can predict the bifurcations
			occurring in the solutions of the 3d elasticity problem
			when the surface tension is increased, leading to a
			localization phenomenon akin to phase separation.
			Comparisons with finite-element simulations reveal that
			the 1d model resolves the interface separating two phases
			accurately, including well into the localized regime, and
			that it has a vastly larger domain of validity than 1d
			model proposed so far.
		\end{abstract}

		
	\end{frontmatter}

\section{Introduction}

\begin{figure}[h!]
	\centering
	\includegraphics[width=10cm]{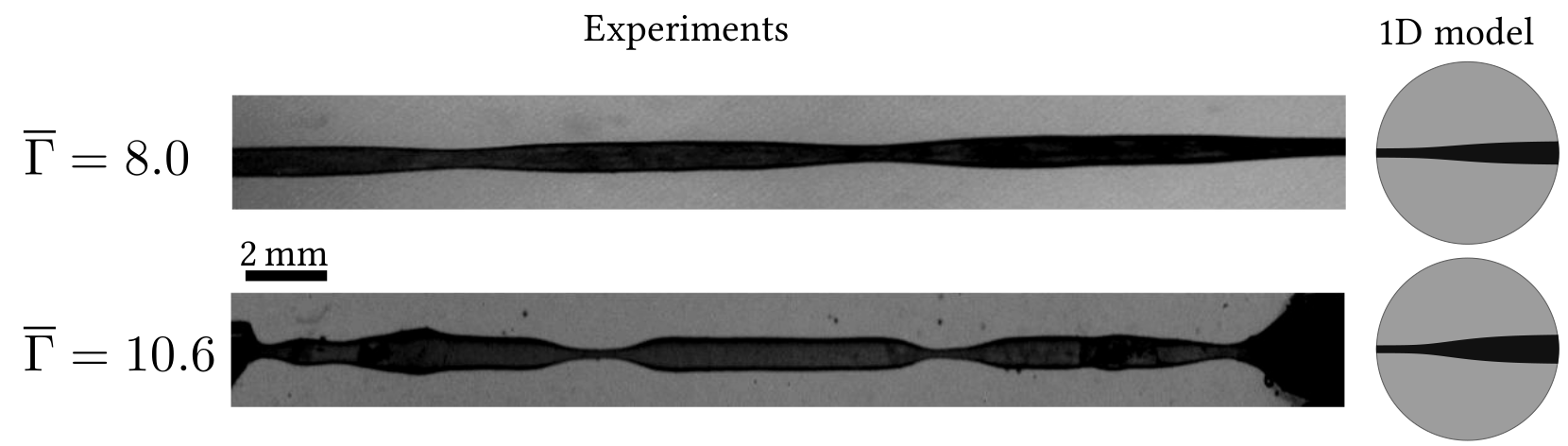}
	\caption{Equilibrium shapes of agar gel filaments with varying
	shear modulus $G$, fixed surface tension $\Gamma$ and undeformed
	radius $\rho$.  The instability is controlled by the dimensionless
	parameter $\overline{\Gamma} = \Gamma/ \, (\rho G)$ corresponding
	to the ratio of the elasto-capillary length to the cylinder's
	radius; the critical value is
	$\overline{\Gamma}_{\mathrm{c}}\approx 5.7$. The aspect ratio of the
	filaments is $\epsilon=\rho/L \approx 0.012$.  The profile of the
	interfaces between regions with constant radius is shown in the
	second column for the same set of parameters, as predicted by 1d
	strain gradient model derived in this paper.  Experimental
	pictures courtesy of Serge Mora.}
	\label{fig:manip}
\end{figure}

We consider a phase separation phenomenon occurring in very soft
elastic cylinders immersed in a fluid.  This phenomenon, first
reported in~\citet{mora2010capillarity}, is shown in
figure~\ref{fig:manip}.  It is driven by surface tension, and is the
elastic analogue of the Rayleigh-Plateau instability in fluids.  It
has been analyzed based on a linear bifurcation
analysis~\citep{mora2010capillarity} and, more recently, on weakly
non-linear bifurcation analyses and on non-linear finite-element
simulations~\citep{taffetani2015beading,xuan2017plateau}.  Here, we
derive a one-dimensional model that captures this phenomenon
accurately, including deeply into the post-bifurcation regime.

Various localization phenomena occurring in non-linear slender
structures have been analyzed recently based on one-dimensional
models, including the necking of hyper-elastic or
bars~\citep{Audoly-Hutchinson-Analysis-of-necking-based-2016}, the
bulging in axisymmetric
balloons~\citep{Lestringant-Audoly-A-diffuse-interface-model-2018}, or
the folding of
tape-springs~\citep{martin2020planar,Brunetti-Favata-EtAl-From-Foppl--von-Karman-2020}. Even though different one-dimensional models have been proposed for
each one of these phenomena, these models are all mathematically
similar. 

\noindent Like
the diffuse-interface model of~\citet{vdwaals94}, they
are built on a strain energy potential made up of a non-convex
function of the elastic strain, plus a regularizing term function of
the gradient of strain---these models will be generically referred to
as \emph{strain gradient models}.
As is well known in the context of one-dimensional solid-solid phase
separation~\citep{Ericksen75,
Carr-Gurtin-EtAl-Structured-phase-transitions-1984}, strain gradient
models can account for localization phenomena: the non-convex energy
tends to produce coexisting phases of constant strain while the
regularizing term penalizes sharp variations of the strain, thereby
accounting for the presence of diffuse interfaces between phases and
for their energy.

One-dimensional (1d) models offer an efficient approach to
localization phenomena and remove the need for complex and
computationally costly simulations, such as the non-linear shell
simulations that are traditionally used for the analysis of tape
springs.  This makes 1d models very attractive.  However these models
have often been derived from \emph{ad hoc} kinematic
assumptions~\citep{xuan2017plateau, martin2020planar}: their accuracy
is not well controlled and their domain of validity is unclear.

This work is part of an ongoing effort from the authors to derive 1d
models for various non-linear slender structures, based on a
systematic two-scales expansion method.  The 1d models obtained in
this way are asymptotically correct in the limit where the
length-scale over which the strain varies is much larger than the
transverse dimension of the structure, although in several cases they
have been found to remain highly accurate in regions of large strain
gradients.  They also tend to be simpler than alternatives based on
\emph{ad hoc} kinematic assumptions, as discussed
elsewhere~\citep{Audoly-Hutchinson-Analysis-of-necking-based-2016,%
Lestringant-Audoly-A-diffuse-interface-model-2018}.  In this paper, we
apply the general reduction method which we proposed
in~\citet{LESTRINGANT2020103730}, to the case of an axisymmetric,
hyper-elastic cylinder with surface tension; doing so, we obtain a
non-linear one-dimensional strain gradient model that captures
elasto-capillary necking.  We compare its predictions to
finite-element simulations of full three-dimensional (3d) model, and
find good agreement, including deep in the post-bifurcation regime
when the interfaces are well localized.  This complements previous
work on a 1d model for elastic necking without surface tension, where
the comparison with the full 3d model was limited to the neighborhood
of the
bifurcation~\citep{Audoly-Hutchinson-Analysis-of-necking-based-2016}.

For cylinders made of very soft solids, the elasto-capillary length,
defined as the ratio of the surface tension over the shear modulus, is
comparable to the cylinder radius.  In this regime, surface tension
competes with elasticity, generating localized patterns akin to those
produced by the Plateau-Rayleigh instability in fluids.  This was
experimentally demonstrated by~\citet{mora2010capillarity}
in filaments of agar gel.  By varying the shear modulus $G$ for a
fixed surface tension $\Gamma$ and a fixed undeformed radius $\rho$,
they showed that the dimensionless parameter $\overline{\Gamma} =
\Gamma/ (\rho\, G)$ determines the width of the interfaces, as
illustrated in Figure~\ref{fig:manip}.  Using a linear stability
analysis they found that the instability occurs when
$\overline{\Gamma}$ exceeds a critical value
$\overline{\Gamma}_{\text{c}}$.  This instability has been further
explored and characterized
by~\citet{taffetani2015beading,taffetani2015elastocapillarity}
using 3d finite elasticity theory combined with linear and weakly nonlinear bifurcation analysis and numerical simulations in the nonlinear regime. Recently,~\citet{xuan2017plateau} revisited the elasto-capillary buckling
as a phase separation process, and used Maxwell's construction to
predict the amplitude of the localization pattern.  Based on a weakly
non-linear analysis, they derived a 1d strain gradient model capturing
the onset of localization, \emph{i.e.}, when
$\overline{\Gamma}-\overline{\Gamma}_{\text{c}} \ll 1$.

In this paper, we extend the work of Xuan \& Biggins and derive a 1d
gradient model that is not limited to the onset of localization.  Our
model allows finite variations of the axial strain across the length
of the cylinder, and does not assume the parameter
$\overline{\Gamma}-\overline{\Gamma}_{\text{c}}$ to be small.  It is
asymptotically justified when the gradient of the axial strain is
small compared to the inverse radius of the cylinder but remains
highly accurate even when this assumption is not satisfied,
\emph{i.e.}, in the presence of the sharp interfaces that arise for
relatively large values of
$\overline{\Gamma}/\overline{\Gamma}_{\text{c}}$.

\section{Problem formulation and summary of the main result}
\label{s:formulation}

\subsection{Finite-elasticity problem in 3d}
\begin{figure}
	\centering
	\includegraphics{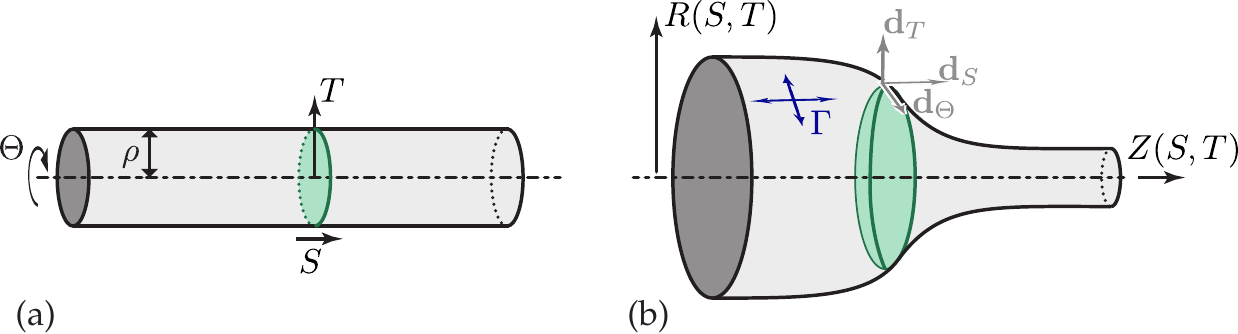}
	\caption{A hyper-elastic cylinder in (a)~reference (undeformed)
	configuration and (b)~actual configuration.}
	\label{fig:3dkinematics}
\end{figure}

We consider a hyper-elastic cylinder with undeformed length $L$ and
undeformed radius $\rho$.  The aspect-ratio $\epsilon=\rho/L$ is
assumed to be small, $\epsilon\ll 1$.  The cylinder deforms in an
axisymmetric way under the combined effect of a traction force applied
at its two ends, and of surface tension of its lateral boundary; the
surface tension is represented in the model by an energy $\Gamma$ per
unit area of the lateral surface, measured in actual configuration.
The cylinder is assumed to be homogeneous, and its elastic
constitutive law to be transversely isotropic (this includes the case
of an isotropic material as a particular case).  With these
assumptions, there exist solutions to the elasticity problem with
surface tension, such that the cross-sections remain planar and
perpendicular to the axis; note that these `fundamental' solutions may
however not be stable.  We use the cylindrical coordinates
$(T,\Theta, S)$ in the reference, undeformed configuration as
Lagragian variables and we denote by $(\mathbf{d}_{S},
\mathbf{d}_{T}(\Theta), \mathbf{d}_{\Theta}(\Theta))$ the
corresponding moving orthonormal frame, as sketched in
Figure~\ref{fig:3dkinematics}(a).
The current position of the point with Lagrangian coordinates
$(T,\Theta,S)$ is $Z (S, T)\, \mathbf{d}_S  + R (S, T)\,
\mathbf{d}_T(\Theta)$, 
as shown in Figure~\ref{fig:3dkinematics}(b).  In axisymmetric
geometry, the deformation gradient writes $\mathbf{F}= Z_{, S}
\,\mathbf{d}_S \otimes \mathbf{d}_S + Z_{, T}\, \mathbf{d}_S \otimes
\mathbf{d}_T + R_{, S} \, \mathbf{d}_T \otimes \mathbf{d}_S + R_{, T}\,
\mathbf{d}_T \otimes \mathbf{d}_T + \frac{R}{T}\, \mathbf{d}_{\Theta}
\otimes \mathbf{d}_{\Theta}$. The Green-Lagrange strain can then be 
calculated as
\begin{equation}
\mathbf{E} = \frac{1}{2}\left(\mathbf{F}^T\cdot \mathbf{F} - \mathbf{I}\right) = \frac{1}{2}\begin{pmatrix}
Z_{,S}^2 + R_{,S}^2 - 1 & Z_{,S}Z_{,T} + R_{,S}R_{,T} & 0\\
Z_{,S}Z_{,T} + R_{,S}R_{,T} & Z_{,T}^2 + R_{,T}^2 - 1 & 0\\
0 & 0 & \left(\left(\frac{R}{T}\right)^2 - 1\right)
\end{pmatrix}_{(S,T,\Theta)},
\label{eq:GLStrain}
\end{equation}
where the matrix notation provides the components of the tensor in the
orthonormal local basis $(\mathbf{d}_{S}, \mathbf{d}_{T}(\Theta),
\mathbf{d}_{\Theta}(\Theta))$, as indicated by the label
$(S,T,\Theta)$ in subscript of the matrix.

A model for the elastic material is specified through a strain energy
potential $w (\mathbf{E})$; we work for the moment with a generic
material model that is transversely symmetric with respect to the axis
$Z$, and compressible.  Later, we will focus on the
special case of an incompressible neo-Hookean material.

The area magnification factor on the lateral surface is the quantity
\begin{equation}
	J = \left.\frac{R}{T} \sqrt{Z_{, S}^2 + R_{, S}^2}
\right|_{T = \rho}\textrm{.}
	\label{eq:JDef}
\end{equation}
The sum of the strain energy and the capillary energy of the system
writes
\begin{equation}
\Phi_{\text{3d}} = \int_0^L \left(\int_0^{\rho} w(\mathbf{E}) 2 \pi T \mathd T + \Gamma 2 \pi \rho J\right) \mathrm{d}S.
\label{eq:NLElasticCylinderEnergy_simple}
\end{equation}
The dimension reduction will be carried out on the potential
$\Phi_{\text{3d}}$: the forces applied on the endpoints of the
cylinder will be introduced later.

\subsection{Dimension reduction strategy}
\begin{figure}
	\centering
	\includegraphics[width=8cm]{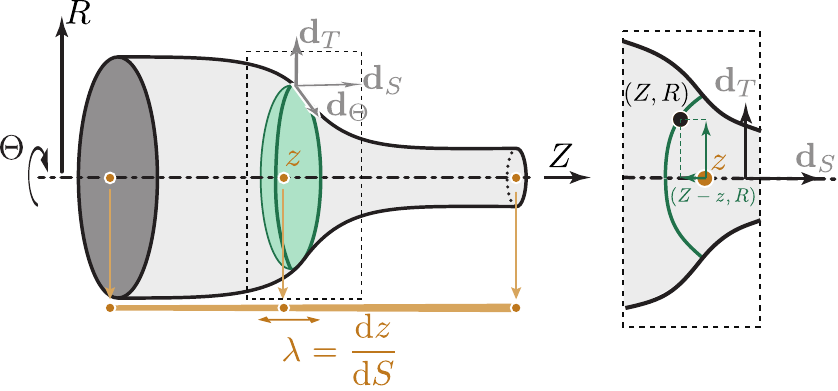}
	\caption{3d kinematics in actual configuration.  The macroscopic
	deformation is captured by the position $z(S)$ along the $Z$-axis
	of the center-of-mass of the cross-section with label $S$, while
	the detailed deformation in the cross-sections is fully described
	by the microscopic displacement $(Z(S,T)-z(S), R(S,T))$ (inset).}
	\label{fig:dimred}
\end{figure}

With a view of conducting a dimension reduction procedure and thus
reducing the kinematics of the system to a macroscopic apparent
deformation, we define the average current coordinate $z(S)$ of the
cross-section whose undeformed coordinate is $S$,
\begin{equation}
z(S) =\langle Z \rangle (S) =\frac{1}{\pi \rho^2} \int_0^\rho Z(S,T) 
\,2 \pi T\, \mathrm{d}T,
\label{eq:averageZ}
\end{equation} 
where $\langle f \rangle$ denotes the average of a quantity $f$ over a
cross-section.  By construction, $z(S)$ is a function of the
longitudinal coordinate $S$ only, see Figure~\ref{fig:dimred}(a); it
will be the main kinematic variable of the 1d model.  We define the
\emph{apparent} axial stretch ratio $\lambda(S)$ as $\lambda (S)=
\frac{\mathd z}{\mathd S}$.  The quantity $\lambda(S)$ with be the
strain measure of our 1d model; it is apparent in the sense that
$z(S)$ is an average position and not the coordinate of any material
point.

We focus on configurations of the cylinders such that typical scale of
variation of $\lambda(S)$ is $L$: $\lambda(S)$ does not
vary significantly on the scale $\rho\ll L$.  In addition we allow
finite variations of the stretch, $\lambda(S)=\mathcal{O}(1)$.  With
the convention that $L=\mathcal{O}(1/\epsilon)$ and
$\rho=\mathcal{O}(1)$, this leads to the scaling assumption
\begin{equation}
	\lambda(S) = \mathcal{O}(1),\quad
	\lambda'(S) = \mathcal{O}(\epsilon),\quad
	\lambda''(S) = \mathcal{O}(\epsilon^{2}),\quad
	\textrm{etc.}
	\label{eq:scalingAssumptions}
\end{equation}
The reduction strategy documented in~\citet{LESTRINGANT2020103730} is
implemented by (\emph{i})~expanding the potential $\Phi_{\text{3d}}$
in~(\ref{eq:NLElasticCylinderEnergy_simple}) in powers of $\epsilon$,
and (\emph{ii})~by relaxing the microscopic displacement
$(Z(S,T)-z(S),R(S,T))$ order by order in $\epsilon$.  By `relaxing the
microscopic displacement', we mean that the functions $Z(S,T)$ and
$R(S,T)$ are solved for in terms of the prescribed macroscopic
variable $z(S)$, from the condition that $\Phi_{\text{3d}}$ is
stationary among all functions $Z(S,T)$ and $R(S,T)$ satisfying the
constraint $\langle Z\rangle(S)=z(S)$ for all $S$, see
equation~(\ref{eq:averageZ}).  Based on this relaxation method, the
microscopic displacement $(Z(S,T),R(S,T))$ is obtained as a function
of the macroscopic displacement $z(S)$, and the potential
$\Phi_{\text{3d}}$ is effectively reduced to a one-dimensional
potential $\Phi_{1d}[\lambda]$.  The dimension reduction strategy is
therefore variational, and \emph{ad hoc} kinematic hypothesis is
involved in the process.

\subsection{Analysis of homogeneous solutions}
Let us analyze homogeneous solutions first.  In view of the assumed
transverse isotropy of the material, solutions with
constant axial stretch ratio $\lambda$ correspond to
\begin{equation}
	Z_{\text{hom}}(S,T)=\lambda\, S + \mathrm{Cte}, \quad
	R_{\text{hom}}(S,T) = \mu T
	\nonumber
\end{equation}
where the transverse stretch ratio $\mu$ is a function of $\lambda$
that captures Poisson's effect in the non-linear setting.  The
associated Green-Lagrange strain is $\mathbf{E}_{\text{hom}}(\lambda,
\mu) = \frac{1}{2}\text{diag}\big(\lambda^2 - 1, \mu^2 - 1, \mu^2 -
1\big)_{(S,T,\Theta)}$ and the area magnification factor is
$J_{\mathrm{hom}}(\lambda,\mu)= \lambda\,\mu$.  We define the strain
energy density of a homogeneous solution as $w_{\text{hom}}(\lambda,
\mu) = w(\mathbf{E}_{\text{hom}}(\lambda, \mu))$.  The associated
Piola-Kirchhoff stress is diagonal and writes
\begin{equation}
\tmmathbf{\Sigma}_{\text{hom}}(\lambda) =
\frac{\partial w}{\partial \mathbf{E}}(\mathbf{E}_{\text{hom}}(\lambda, \mu_\lambda))= \text{diag}\left(\Sigma_{\text{hom}}^\parallel(\lambda, \mu), \Sigma_{\text{hom}}^\perp(\lambda, \mu), \Sigma_{\text{hom}}^\perp(\lambda, \mu) \right),
\label{eq:sigmahom}
\end{equation}
where $\Sigma_{\text{hom}}^{\parallel}(\lambda, \mu)= \frac{1}{\lambda}\frac{\partial
w_{\text{hom}}}{\partial \lambda}(\lambda, \mu)$ and 
$\Sigma_{\text{hom}}^{\perp}(\lambda, \mu)= \frac{1}{2 \, 
\mu}\frac{\partial w_{\text{hom}}}{\partial \mu}(\lambda, \mu)$ are 
the longitudinal and transverse Piola-Kirchhoff stress in the 
homogeneous solution, respectively.

The quantity $\mu$ is found by setting to zero the first variation of the strain energy
$\Phi_{\text{3d}}$ in~(\ref{eq:NLElasticCylinderEnergy_simple}) with
respect to the transverse strain $\mu$.  This yields an implicit
equation for $\mu$ expressing the equilibrium in the transverse
direction,
\begin{equation}
\frac{\Gamma \lambda}{\rho} + \mu \, \Sigma_{\text{hom}}^{\perp}(\lambda, \mu) = 0.
\label{eq:nlCylinderEquil_simple_announce}
\end{equation}
The solution $\mu$ of this equation is denoted as $\mu_\lambda$.

\subsection{Main result}

The outcome of the reduction method is as follows.  The full 3d
energy~(\ref{eq:NLElasticCylinderEnergy_simple}) reduces to a 1d
potential that depends on the macroscopic strain $\lambda$ only,
\emph{i.e.}, $\Phi_{\text{3d}} \approx \Phi_{\text{1d}}[\lambda]$ where
\begin{subequations}
	\label{eq:1dmodelAnnounce}
\begin{equation}
	\Phi_{\text{1d}}[\lambda] = \int_0^L \left( W_{\text{hom}} (\lambda (S))+\frac{1}{2} \mymult  \mathrm{B}_{\lambda (S)} \mymult \lambda^{\prime \mymult 2} (S)\right)
	\mathd S + [\mathrm{C}_{\lambda (S)} \mymult
	\lambda' (S)]_0^L + \mathcal{O}(L\,\epsilon^3).
	\label{eq:RelaxedEnergy_announce}
\end{equation}
This model is asymptotically exact to order $\epsilon^{2}$.  The
dependence of $\Phi_{\text{1d}}[\lambda]$ on $\lambda$ is non-linear,
as the model accounts for finite stretch ratios and for the (yet
unspecified) non-linear material law.  The quantity
\begin{equation}
	W_{\text{hom}}(\lambda) = \pi\rho^{2}\, w_{\mathrm{hom}}(\lambda,\mu_{\lambda})
	+\Gamma\,2\pi\rho\,\lambda\,\mu_{\lambda}
	\label{eq:Whom}
\end{equation}
defines the 1d model at order $\epsilon^{0}$: it is
a non-linear potential that characterizes homogeneous solutions with
uniform axial stretch ratio $\lambda$.  For the problem at hand and
for other problems involving localization, the potential
$W_{\text{hom}}(\lambda)$ may be non-convex for some values of the
parameters.  

The coefficients $\mathrm{B}_{\lambda}$ and $\mathrm{C}_{\lambda}$ in
equation~(\ref{eq:RelaxedEnergy_announce}) are found in the
forthcoming section as
\begin{equation}
\frac{1}{2}\mathrm{B}_{\lambda} =  \pi \rho^4 \frac{ (\nabla \mu_{\lambda})^2}{4} 
\left( \Sigma_{\text{hom}}^{\parallel}(\lambda, \mu_\lambda) + 5
\frac{\Gamma}{\rho}  \frac{\mu_{\lambda}}{\lambda} \right) \quad \text{and} \quad  \mathrm{C}_{\lambda}
= - \pi \, \rho^3 
	\frac{\Gamma}{2}\,  \frac{\mu_{\lambda}^2}{\lambda} \,\nabla \mu_{\lambda},
\label{eq:gradientmoduli_announce}
\end{equation}
where the $\nabla$ notation stands for derivatives with respect to the 
axial stretch $\lambda$,
\begin{equation}
	\nabla \mu_{\lambda} = \frac{\mathrm{d} \mu_\lambda}{\mathrm{d}\lambda}.
	\label{eq:nablaNotation}
\end{equation}
\end{subequations}

One can derive the equations of equilibrium of the 1d model by applying
the calculus of variations to the potential $\Phi_{\text{1d}}$ in
equation~(\ref{eq:RelaxedEnergy_announce}),
see~\S\ref{s:dimred}\ref{s:order2}.  The resulting set of ordinary
differential equations are much easier to solve numerically than the
original finite elasticity model; they are also well suited for
bifurcation and stability analyses, see
\S\ref{s:results}\ref{ssec:taffetaniCompare}
or~\citet{Lestringant-Audoly-A-diffuse-interface-model-2018} .  We will
demonstrate the accuracy of this model in section~\ref{s:results} by
comparing its predictions to numerical simulations of the full
model~(\ref{eq:NLElasticCylinderEnergy_simple}) obtained by the finite
element method.

\section{Dimension reduction}
\label{s:dimred}

In the following, the primes are reserved for the derivation with
respect to the longitudinal coordinate $S$ while the derivation with
respect to the radial coordinate is represented by $\partial_T$.  For
a function $f(S,T)$, this writes
\[ f' = \frac{\partial f}{\partial S}, \quad \partial_T f = \frac{\partial f}{\partial T}.\]

We consider a prescribed distribution of the macroscopic axial strain
$\lambda(S)$ with $0\leq S \leq L$, such that the scaling assumptions
in~(\ref{eq:scalingAssumptions}) hold.  We seek a current 
configuration of the cylinder compatible with the prescribed 
macroscopic strain $\lambda(S)$ in the form
\begin{subequations}
	\label{eq:microscopic_disp}
\begin{align}
	Z(S,T) &= \int_{0}^{S} \lambda(\tilde{S})\, \mathrm{d}\tilde{S} + \frac{1}{2} 
	\mathcal{A}(S) \left(T^2 - 
	\frac{\rho^2}{2}\right)+\mathcal{O}(\epsilon^3)
	\label{eq:microscopic_disp-Z}\\
	R(S,T) &=  \mu(S) \,T + \mathcal{B}(S,T)+\mathcal{O}(\epsilon^4),
	\label{eq:microscopic_disp-R}\\
	\intertext{where}
	&\mathcal{A}(S) = \mathcal{O}(\epsilon), \quad \mathcal{B}(S,T)=\mathcal{O}(\epsilon^2), \quad \frac{\mathrm{d}}{\mathrm{d}S} = \mathcal{O}(\epsilon).
	\label{eq:ABscalingAssumptions}
\end{align}
\end{subequations}
Note that this expansion~(\ref{eq:microscopic_disp}) is such that
$\langle Z\rangle(S)=\int\lambda \,\mathrm{d}S=z(S)$, so that the
constraint~(\ref{eq:averageZ}) is automatically satisfied.  The
correction $\mathcal{A}(S)$ to the longitudinal displacement curves
the initially planar cross-sections into paraboloids having radial
curvature $\mathcal{A}(S)$.  As we will show, the correction
$\mathcal{B}(S,T)$ to the radial displacement plays no role in the
asymptotic expansion of the energy up to order $L\,\epsilon^2$: to
determine this quantity, one would need to push the expansion to a
higher order.

We have argued in section~\ref{s:formulation} that our reduction
method is asymptotically correct, and free of any \emph{ad hoc}
assumption, and it may seem paradoxical that it starts with
equation~(\ref{eq:microscopic_disp}) which looks like a kinematic
assumption.  It is actually possible to justify the form of the
expansion~(\ref{eq:microscopic_disp}) by asymptotic analysis, by
applying the systematic reduction method
from~\citet{LESTRINGANT2020103730}.  This full derivation
works along exactly the same lines as the worked examples presented
in~\citet{LESTRINGANT2020103730}; it is straightforward but somewhat
tedious.  It is not included here, and we use the
form~(\ref{eq:microscopic_disp}) as a starting point to keep the
presentation concise.  It can be checked at the end of the calculation
that our solution satisfies all the equations of equilibrium to the appropriate order in
$\epsilon$.

The dimension reduction method is variational: among the fields of the
form~(\ref{eq:microscopic_disp}), we seek the one that minimizes the
strain energy $\Phi [\lambda, \mu, \mathcal{A}, \mathcal{B}]$ obtained
by inserting the displacement~(\ref{eq:microscopic_disp}) into the 3d
strain energy~(\ref{eq:NLElasticCylinderEnergy_simple}),
\begin{subequations}
	\label{eq:insertTrialDisplacementIntoEnergy}
\begin{equation}
\Phi [\lambda, \mu,  \mathcal{A}, \mathcal{B}] =
\int_0^L \left(\int_0^{\rho} w 
\big(\mathbf{E}^{\dag} \big) 2 \pi 
T\, \mathd T
+ 2 \pi \rho \,\Gamma  J^{\dag}\right)\mathrm{d}S.
\label{eq:NLElasticCylinderEnergy}
\end{equation}
Here, the strain is obtained as
\begin{equation}
\mathbf{E}^{\dag}
= 
\frac{1}{2}
\left(\begin{smallmatrix}
\left(\lambda +  \frac{\mathcal{A}'}{2}\zOfT(T)\right)^2 + (\mu'T + \mathcal{B}')^2 - 1 & \text{sym} & 0 \\
\left(\lambda + \frac{\mathcal{A}'}{2}\zOfT(T)\right) T \mathcal{A} + (\mu'T + \mathcal{B}')
(\mu + \partial_T \mathcal{B})  &  (T \mathcal{A})^2 +(\mu + \partial_T \mathcal{B})^2 - 1 & 0\\
0 & 0 &   \left(\mu+\frac{\mathcal{B}}{T}\right)^2 - 1 
\end{smallmatrix}\right)
_{(S,T,\Theta)}
\label{eq:GLstrain_detailled}
\end{equation}
where the `$\mathrm{sym}$' entry is repeated from across the diagonal 
by symmetry and $\zOfT(T) = T^2 - \frac{\rho^2}{2}$.  
The area
magnification factor writes
\begin{equation}
J^{\dag} = 
\left(\mu+\frac{\mathcal{B}(S,\rho)}{\rho}\right) 
\,\sqrt{\left(\lambda + \frac{\mathcal{A}'\,\rho^{2}}{4}\right)^2+ \left(\mu'\rho+\mathcal{B}'(S,\rho)\right)^2}
\textrm{.}
\label{eq:Jnew}
\end{equation}
\end{subequations}
The final 1d energy
$\Phi_{\text{1d}} [\lambda]$ is obtained by minimizing the potential
energy~(\ref{eq:insertTrialDisplacementIntoEnergy}) with respect to
the functions $\mu$, $\mathcal{A}$ and $\mathcal{B}$ order by order in
$\epsilon$.

\subsection{Order $\epsilon^{0}$: non-convex bar model}
\label{s:order0}

At order $0$ in $\epsilon$, the axial gradients such as $\lambda'$ and
$\mu'$, as well as the coefficients involving $\mathcal{A}(S)$ and
$\mathcal{B}(S,T)$ do not contribute to the
strain~(\ref{eq:GLstrain_detailled}--\ref{eq:Jnew}), see the scaling assumptions
in~(\ref{eq:scalingAssumptions}) and~(\ref{eq:ABscalingAssumptions}).
The potential energy~(\ref{eq:NLElasticCylinderEnergy}) thus writes
\begin{equation}
\Phi
[\lambda, \mu,\mathcal{A}, \mathcal{B}] = \int_0^L\left( \pi 
\rho^2 w_{\text{hom}}(\lambda(S), \mu(S)) + 2 \pi \rho \mymult  
\Gamma\lambda(S) \mymult \mu(S)\right)\mathrm{d}S + \mathcal{O}(L\,\epsilon)
\label{eq:order0Energy}
\end{equation}
where $w_{\mathrm{hom}}$ and $\mathbf{E_{\mathrm{hom}}}$ have been
defined in~(\ref{eq:nlCylinderEquil_simple_announce}) and immediately
before~(\ref{eq:nlCylinderEquil_simple_announce}), respectively.

For a given distribution $\lambda$ of apparent stretch ratio, the
minimization of the potential energy~(\ref{eq:order0Energy}) with
respect to $\mu(S)$ yields an implicit equation for the transverse
stretch $\mu(S)$ as
\begin{equation}
\frac{\Gamma \lambda}{\rho} + \mu(S)  \,\Sigma_{\text{hom}}^{\perp}(\lambda(S), \mu(S)) = 0.
\label{eq:nlCylinderEquil_simple}
\end{equation}
Equation~(\ref{eq:nlCylinderEquil_simple}) expresses the fact that the
uniform transverse stress inside the cylinder is at equilibrium with
the surface tension at the boundary; it is exactly the same relation
as obtained earlier in
equation~(\ref{eq:nlCylinderEquil_simple_announce}) when we analyzed
homogeneous solutions.  Its solution $\mu$ is a function of $\lambda$
that is denoted as $\mu_{\lambda}$, where the mapping
$\lambda\mapsto\mu_{\lambda}$ is the catalog of homogeneous solutions
labelled by the (uniform) axial stretch $\lambda$.  Stated
differently, we have shown that the transverse stretch must match, at
order $\epsilon^{0}$, the transverse stretch $\mu_{\lambda(S)}$
predicted by the analysis of homogeneous solutions, as if the stretch
were everywhere equal to the local stretch ratio $\lambda(S)$,
\begin{equation}
	\mu(S) = \mu_{\lambda(S)}
	\textrm{.}
	\label{eq:optimalMu}
\end{equation}
Inserting this into~(\ref{eq:order0Energy}), we obtain
\begin{equation}
\Phi_{\text{1d}}\left[\lambda\right] = \int_0^L W_{\text{hom}} 
(\lambda (S))\, \mathrm{d}S+\mathcal{O}(L\,\epsilon),
\label{eq:1Dmodel0}
\end{equation}
where $W_{\mathrm{hom}}(\lambda)$ has been defined in~(\ref{eq:Whom}).
This is in agreement with the 1d model introduced
in~\citet{xuan2017plateau}.

The potential~(\ref{eq:1Dmodel0}) defines a non-linear bar model; no
dependence on the strain gradient is present at this order.  The
potential $W_{\mathrm{hom}}$ is non-convex when the surface tension
$\Gamma$ exceeds a critical value~\citep{xuan2017plateau}: in such
circumstances, it admits non-smooth solutions made of two or more
phases of distinct stretch ratios $\lambda$, separated by sharp
interfaces.  This non-regularized model is therefore akin to
Ericksen's bar model introduced in the context of
necking~\citep{Ericksen75}.  Both the stretch ratios in the two phases
and the overall transformation stretch can be predicted using
Maxwell's construction~\citep{MaxwellNature}, as observed
in~\citet{xuan2017plateau}.  For example, for an incompressible
neo-Hookean material with a shear modulus $G$, the homogeneous
solution is $\mu_\lambda = 1/\sqrt{\lambda}$ and the 1d energy density
defined in~(\ref{eq:Whom}) and~(\ref{eq:1Dmodel0}) writes, see
equation~(\ref{eq:1d-nHincompressible-Whom}) in
appendix~\ref{a:incompressibleCase},
\begin{equation}
W_{\text{hom}}(\lambda) = \frac{\mathpi \, \rho^2 \, G}{2}
   \, \left( \lambda^2 + \frac{2}{\lambda} + 4 \, \overline{\Gamma}
   \, \lambda^{1 / 2} \right)
\qquad\textrm{(incompressible neo-Hookean material)}
\label{eq:order0_biggins}
\end{equation}
where the dimensionless parameter $\overline{\Gamma} = \Gamma/(G
\rho)$ measures the ratio of the elasto-capillary length to the
diameter of the cylinder.  The critical value for the surface tension
associated with a loss of convexity for $W_{\text{hom}}$ is then
$\overline{\Gamma}_\text{c} = \sqrt{32}$, as found
in~\citet{xuan2017plateau}.

The 1d bar model in~(\ref{eq:order0Energy}) captures phase
separation~\citep{xuan2017plateau} but it cannot predict the shape of
interfaces between the phases as the gradient $\lambda'(S)$ has been
neglected so far.  In addition, the 1d bar model
in~(\ref{eq:order0Energy}) does not capture finite-size effects, such
the dependence on the aspect-ratio $L/\rho$ of the critical surface
tension $\Gamma$ that makes the homogeneous solution unstable.  With
the aim to overcome these limitations, we push the asymptotic
expansion further in the next section, and derive a regularized 1d
model that accounts for the strain energy associated with the strain
gradient $\lambda'$.

For further reference we note the Piola-Kirchhoff stress in the
homogeneous configuration as
\begin{equation}
	\tmmathbf{\Sigma}_{\text{hom}}(\lambda) =
\frac{\partial w}{\partial \mathbf{E}}(\mathbf{E}_{\text{hom}}(\lambda, \mu_\lambda))= \text{diag}\left(\Sigma_{\text{hom}}^\parallel(\lambda), \Sigma_{\text{hom}}^\perp(\lambda), \Sigma_{\text{hom}}^\perp(\lambda) \right)
\textrm{,}
\label{eq:Sigma-hom}
\end{equation}
where, as earlier in equation~(\ref{eq:sigmahom}),
$\Sigma_{\text{hom}}^\parallel(\lambda) =
\Sigma_{\text{hom}}^\parallel(\lambda,\mu_\lambda)$ and
$\Sigma_{\text{hom}}^\perp(\lambda) =
\Sigma_{\text{hom}}^\perp(\lambda,\mu_\lambda)$.  We also define the
tangent stiffness in the homogeneous configuration
$\tmmathbf{K}_{\text{hom}}(\lambda)$ as the operator entering in the
expansion of the strain energy about $\mathbf{E}_{\text{hom}}$,
\begin{subequations}
	\label{eq:w-expansion-to-second-order}
	\begin{equation}
		w(\mathbf{E}_{\text{hom}} + \mathbf{\hat{E}}) = w(\mathbf{E}_{\text{hom}}) + \tmmathbf{\Sigma}_{\text{hom}}(\lambda) : \mathbf{\hat{E}} + \frac{1}{2}\mathbf{\hat{E}}:\tmmathbf{K}_{\text{hom}}(\lambda):\mathbf{\hat{E}}
		+\cdots
		\label{eq:w-expansion-to-second-order-A}
	\end{equation}
	Here, the double dot notation stands for a double contraction, as in
$\tmmathbf{\Sigma}_{\text{hom}} : \mathbf{\hat{E}} =
\sum_{i,j=1}^{3}\Sigma_{\mathrm{hom},ij}\,\hat{E}_{ij}$.
	We finally define the (scalar) tangent shear modulus
	$K_{\text{hom}}^{STST}(\lambda)$ in the homogeneous configuration
	which for transversely symmetric materials is such that
\begin{equation}
	\tmmathbf{K}_{\text{hom}}(\lambda):\left(\mathbf{d}_S\otimes\mathbf{d}_T+\mathbf{d}_T\otimes\mathbf{d}_S\right) = 2 \,K_{\text{hom}}^{STST}(\lambda)\left(\mathbf{d}_S\otimes\mathbf{d}_T+\mathbf{d}_T\otimes\mathbf{d}_S\right).
	\label{eq:w-expansion-to-second-order-B}
\end{equation}
\end{subequations}

\subsection{Optimal correction}
\label{s:optcorrec}

Here, we push the expansion to order $\epsilon^{2}$, and calculate the
optimal values of the quantities $\mathcal{A}(S)$ and
$\mathcal{B}(S,T)$ in~(\ref{eq:microscopic_disp}) that minimize the
energy~(\ref{eq:NLElasticCylinderEnergy}).

With a view of expanding $\Phi
[\lambda,\mu_\lambda,\mathcal{A},\mathcal{B}]$ in powers of the
successive gradients of $\lambda$ up to order~$\epsilon^{2}$, we start
by expanding the Green-Lagrange deformation
gradient~(\ref{eq:GLstrain_detailled}):
\begin{subequations}
	\label{eq:GLstrainExpansion}
\begin{equation}
\mathbf{E}^{\dag} = \mathbf{E}_{\text{hom}}(\lambda,\mu_\lambda) + \mathbf{E}_1 + \mathbf{E}_2+ \mathcal{O}(\epsilon^3),
\end{equation}
where
\begin{align}
	\mathbf{E}_1 & = E_{1}^{ST}\,
	\left(
	\mathbf{d}_{S}\otimes \mathbf{d}_{T} + 
	\mathbf{d}_{T}\otimes \mathbf{d}_{S}
	\right),
	\qquad
	E_{1}^{ST}=\frac{(\mathcal{A} \lambda +\nabla \mu_\lambda 
	\lambda' \mu)\,T}{2}
\label{eq:GLstrainExpansion-E1}
\\
	\mathbf{E}_2 & = \frac{1}{2}\begin{pmatrix}
(\nabla \mu_\lambda \lambda')^2 T^2 + \mathcal{A}'\lambda \,\zOfT(T) & 0  & 0\\
0& \mathcal{A}^2 T^2 + 2 \mu_\lambda \partial_T \mathcal{B} & 0 \\
0 & 0 &2 \mu_\lambda \frac{\mathcal{B}}{T}
\end{pmatrix}_{(S,T,\Theta)}
\end{align}
\end{subequations}
The radial stretch $\mu(S) = \mu_{\lambda(S)}$ imposed by the
optimality condition at order $\epsilon^{0}$ has been used,
see~(\ref{eq:optimalMu}).  In particular the derivative $\mu'(S)$ has
been calculated as
\begin{equation}
	\mu'(S) = \frac{\mathrm{d}[\mu_{\lambda(S)}]}{\mathrm{d}S} = 
	\frac{\mathrm{d}\mu_{\lambda}}{\mathrm{d}\lambda}\,\frac{\mathrm{d}\lambda}{\mathrm{d}S}
	= \big[\nabla \mu_{\lambda}\big]_{\lambda(S)}\,\lambda'(S),
\end{equation}
using the $\nabla$ notation from~(\ref{eq:nablaNotation}).
The expression of $J^{\dag}$ in~(\ref{eq:Jnew}) is expanded similarly
to order $\epsilon^2$,
\begin{equation}
	\begin{split}
		J^{\dag} & 
		=\left(\mu_{\lambda}+\frac{\mathcal{B}(S,\rho)}{\rho}\right)\,\sqrt{\left(\lambda+ 
		\frac{\mathcal{A}'\,\rho^{2}}{4}\right)^{2} + \left(\nabla 
		\mu_\lambda \lambda' \rho + \mathcal{B}'(S,\rho)\right)^2}  \\
		& = \mu_\lambda \lambda + \mu_\lambda \frac{(\nabla \mu_\lambda)^2 \lambda^{\prime \, 2}\rho^2 + \mathcal{A}'\lambda \, \frac{\rho^2}{2}}{2 \lambda}
		+ \frac{\lambda}{\rho}\,\mathcal{B}(S,\rho)
		+ \mathcal{O}(\epsilon^3).
	\end{split}
\label{eq:JExpansion}
\end{equation}

The expansion of the energy~(\ref{eq:NLElasticCylinderEnergy}) is 
then found as
\begin{subequations}
	\label{eq:enerExpand}
	\begin{equation}
		\Phi[\lambda,\mu_\lambda, \mathcal{A},\mathcal{B}] = \int_0^L 
		W_{\text{hom}}(\lambda(S))\mathrm{d}S + \Phi_{1} + \Phi_{2} + 
		\mathcal{O}(L\,\epsilon^3),
	\end{equation}
	where, after using the special form of $\mathbf{E}_{1}$
	from~(\ref{eq:GLstrainExpansion-E1}) as well as the
	identity~(\ref{eq:w-expansion-to-second-order-B}),
	\begin{align}
		\Phi_{1} & = \int_0^L\left(\int_0^\rho\tmmathbf{\Sigma}_{\text{hom}} : \mathbf{E}_1 \, 2 \pi T\mathrm{d}T\right)\mathrm{d}S\\
		\Phi_{2} & = \int_0^L\Bigg[\int_0^\rho\left(\tmmathbf{\Sigma}_{\text{hom}} : \mathbf{E}_2 + 
		2 K_{\text{hom}}^{STST}\, (\mathrm{E}_{1}^{ST})^2\right)2 \pi T \mathrm{d}T\cdots \nonumber\\
		&\hspace{2cm}
		{}
		 +  
		2 \pi \rho \Gamma \frac{(\nabla \mu_\lambda)^2 
		\lambda^{\prime \, 2}\rho^2 + \mathcal{A}'\lambda\, 
		\frac{\rho^2}{2}}{2\lambda}\mu_\lambda
		+2\pi\,\Gamma\,\lambda\,\mathcal{B}(S,\rho)\Bigg]\mathrm{d}S
		\label{eq:enerExpand-Phi2}
	\end{align}
\end{subequations}
Unnecessary
function arguments have been omitted for the sake of legibility:
quantities in the integrands such as $\tmmathbf{\Sigma}_{\text{hom}} $
and $K_{\text{hom}}^{STST}$ are all evaluated implicitly with $\lambda
= \lambda(S)$.

In the expansion~(\ref{eq:enerExpand}), the term $\Phi_1$ of order
$\epsilon$ is zero because $\tmmathbf{\Sigma}_{\text{hom}}$ is
diagonal and therefore orthogonal to $\mathbf{E}_1$,
\begin{equation}
	\Phi_1=0\textrm{.}
\end{equation}  

The contributions to $\Phi_{2}$ that depend on the radial correction
$\mathcal{B}$ come from $\tmmathbf{\Sigma}_{\text{hom}}(\lambda) :
\mathbf{E}_2$ on the one hand, see~(\ref{eq:GLstrainExpansion}), and
from last term in the integrand in~(\ref{eq:enerExpand-Phi2}) on the
other hand.  These contributions sum up to
\begin{multline}
	\int_0^L
	\left(\mu_{\lambda(S)} \, \Sigma_{\text{hom}}^\perp(\lambda(S))
	\int_0^\rho \left(\partial_T \mathcal{B} +
	\frac{\mathcal{B}}{T}\right) 2 \pi T \mathrm{d}T
	+
	2\pi\,\Gamma\,\lambda\,\mathcal{B}(S,\rho)
	\right) \mathrm{d}S \cdots \\
	{} =
	2 \pi\, \int_0^L
	\left( \mu_{\lambda(S)} \, \Sigma_{\text{hom}}^\perp(\lambda(S))
	\int_0^\rho \frac{\mathrm{d}[\mathcal{B}\,T]}{\mathrm{d}T}\, \mathrm{d}T
	+
	\Gamma\,\lambda\,\mathcal{B}(S,\rho)
	\right) \mathrm{d}S\cdots \\
	{} = 
	2 \pi\, \rho\int_0^L
	\left(\mu_{\lambda(S)} \, \Sigma_{\text{hom}}^\perp(\lambda(S))
	+
	\Gamma\,\frac{\lambda}{\rho}
	\right) \mathcal{B}(S,\rho)\,\mathrm{d}S
	\nonumber
\end{multline}
This quantity is zero by the transverse equilibrium
condition~(\ref{eq:nlCylinderEquil_simple}): the correction
$\mathcal{B}$ to the displacement does not enter in the energy at this
order.

Expanding first term in the integrand of $\Phi_2$ and ignoring the
$\mathcal{B}$ term which has already been processed, we find
\begin{align*}
\int_0^\rho\tmmathbf{\Sigma}_{\text{hom}} : \mathbf{E}_2 \,2 \pi T \mathrm{d}T & = \int_0^\rho\left(\Sigma_{\text{hom}}^\parallel\left((\nabla \mu_\lambda \lambda')^2 T^2 + \mathcal{A}'\lambda (T^2 - \frac{\rho^2}{2}) \right) + \Sigma_{\text{hom}}^\perp \mathcal{A}^2 T^2\right) \pi  T\mathrm{d}T \\
& = \frac{\pi \rho^4}{4}\left( \Sigma_{\text{hom}}^\parallel (\nabla \mu_\lambda \lambda')^2 -\frac{\Gamma}{\rho}\frac{\lambda}{\mu_\lambda}  \mathcal{A}^2 \right),
\end{align*}
where we have used the equilibrium
equation~(\ref{eq:nlCylinderEquil_simple}) to eliminate
$\Sigma_{\text{hom}}^\perp$.  The first term in the second line of $\Phi_2$ can be
integrated by parts,
\begin{multline*}
\int_0^L  2 \pi \rho \Gamma \frac{(\nabla \mu_\lambda)^2 
\lambda^{\prime \, 2}\rho^2 + \mathcal{A}'\lambda 
\,\frac{\rho^2}{2}}{2\lambda}\mu_\lambda \,\mathrm{d}S = \\
\int_0^L\frac{\pi \,\rho^4}{4} \frac{\Gamma}{\rho}\left(4\frac{ \mu_\lambda}{\lambda}(\nabla \mu_\lambda)^2 \lambda^{\prime \, 2} - 2 \mathcal{A}\nabla\mu_\lambda \lambda'\right)\mathrm{d}S
+\left[\pi \rho^4 \frac{\Gamma}{\rho} \frac{\mathcal{A} \mu_\lambda}{2}\right]_0^L.
\end{multline*}
Lastly, we integrate the second term in the first line of $\Phi_{2}$
with respect to the radial coordinate as
\begin{align*}
	\begin{split}
		\int_0^\rho2\,(\mathrm{E}_{1}^{ST})^2 
		\,K_{\text{hom}}^{STST}\,2 \pi T \mathrm{d}T &= 
		K_{\text{hom}}^{STST} \left(\mathcal{A} \lambda +\nabla 
		\mu_\lambda \lambda' \mu\right)^2\,\int_0^\rho \pi\, T^3\, \mathrm{d}T\\
		&= \frac{\pi \rho^4}{4}\, K_{\text{hom}}^{STST} \left(\mathcal{A} \lambda +\nabla \mu_\lambda \lambda' \mu\right)^2. 
	\end{split}
\end{align*}

Inserting all this into~(\ref{eq:enerExpand}), we find
\begin{multline}
	\Phi[\lambda, \mu_\lambda, \mathcal{A}, 0]  =  \int_0^L W_{\text{hom}}(\lambda(S))\mathrm{d}S  +  \int_0^L \pi \frac{\rho^4}{4} (\nabla \mu_\lambda \lambda')^2\left(\Sigma_{\text{hom}}^\parallel +  4\frac{ \Gamma \mu_\lambda}{\rho \lambda}\right)\mathrm{d}S+ \left[\pi \rho^4 \frac{\Gamma}{\rho} \frac{\mathcal{A} \mu_\lambda}{2}\right]_0^L\\
	+ \int_0^L  \frac{\pi\,\rho^4}{4}
	\Bigg(
	\underbrace{
	K_{\text{hom}}^{STST}\left( \mathcal{A} \lambda + \mu_\lambda\,\nabla 
	\mu_\lambda \lambda' \right)^2 - 
	\frac{2\,\Gamma}{\rho\,\mu_{\lambda}}\left(\frac{1}{2}\, 
	\mathcal{A}^2\,\lambda + 
	\mathcal{A}\,\mu_{\lambda}\,\nabla\mu_\lambda\, \lambda'\right) 
	}_{\mathcal{F}(\mathcal{A})}
	\Bigg)\mathrm{d}S.
\label{eq:energyexpanded2}
\end{multline}
The energy~(\ref{eq:energyexpanded2}) is quadratic with respect to the
curvature $\mathcal{A}(S)$.  It is also convex, \emph{i.e.}, the
coefficient of the $\mathcal{A}^{2}(S)$ term is positive,
\begin{equation}
	\lambda\,K_{\mathrm{hom}}^{{STST}}-\frac{\Gamma}{\rho\,\mu_{\lambda}}>0
	\textrm{,}
	\label{eq:A-stability}
\end{equation}
as shown in the appendix in the case of an incompressible material,
see equation~(\ref{eq:KSTST-incompr}).  The following stationarity
condition therefore warrants that the
energy~(\ref{eq:energyexpanded2}) is minimum with respect to
$\mathcal{A}(S)$,
\begin{equation}
	0=\frac{\partial \mathcal{F}}{\partial \mathcal{A}} = 2\,\left( 
	\mathcal{A} \lambda + \nabla \mu_\lambda \lambda' 
	\mu\right)\left(\lambda \,K_{\text{hom}}^{STST}
	-\frac{\Gamma}{\rho\,\mu_{\lambda}}\right)
	\textrm{.}
	\label{eq:optimalityConditionForA}
\end{equation}
The solution is
\begin{equation}
	\mathcal{A}(S) = -\frac{\mu_{\lambda(S)}}{\lambda(S)} \,\big[\nabla 
\mu_\lambda\big]_{\lambda(S)} \lambda'(S)
	\label{eq:Asolution}
\end{equation}
This expression of the curvature $\mathcal{A}(S)$ of the
cross-sections agrees with expressions derived in the literature in
the absence of surface tension, see equation 2.26b
from~\citet{Audoly-Hutchinson-Analysis-of-necking-based-2016}.  The
curvature of the cross-section is such that the first-order correction
to the strain is zero, $\mathbf{E}_{1}= \mathbf{0}$, as revealed by a
comparison of equations~(\ref{eq:GLstrainExpansion-E1})
and~(\ref{eq:optimalityConditionForA}).  This is a nice consequence of
the fact that our derivation is asymptotically correct: when \emph{ad
hoc} kinematic assumptions are used, the curvature of the
cross-sections is typically overlooked, $\mathcal{A}(S)=0$, and a
spurious shear strain $\mathbf{E}_{1}\neq \mathbf{0}$ is obtained; the
associated spurious shear stress on the lateral boundaries makes it
impossible to satisfy the equilibrium, as discussed
in~\citet{Audoly-Hutchinson-Analysis-of-necking-based-2016}.

\subsection{Higher-order bar model}
\label{s:order2}

When the solution for $\mathcal{A}(S)$
from~(\ref{eq:Asolution}) is inserted into the energy
expansion~(\ref{eq:energyexpanded2}), the term proportional to
$K_{\text{hom}}^{STST}$ cancels and the second line
in~(\ref{eq:energyexpanded2}) boils down to $\int_0^L 
\frac{\pi\,\rho^4}{4} \frac{ \Gamma \mu_\lambda}{\rho \lambda}(\nabla
\mu_\lambda \lambda')^2\mathrm{d}S$.  After some algebra, one obtains 
the final expression for the 1d model capturing the gradient effect, 
$\Phi[\lambda, \mu_\lambda, \mathcal{A}, 0] = 
\Phi_{\text{1d}}[\lambda]$, as
\begin{subequations}
	\label{eq:full1dModelAsDerived}
\begin{equation}
\Phi_{\text{1d}}[\lambda] = \int_0^L \left( W_{\text{hom}} (\lambda (S))+\frac{1}{2} \mymult \mathrm{B}_{\lambda (S)}  \mymult \lambda^{\prime \mymult 2} (S)\right)
\mathd S + \Big[\mathrm{C}_{\lambda (S)} \mymult
\lambda' (S)\Big]_0^L +   \mathcal{O}(L\,\epsilon^3),
\label{eq:RelaxedEnergy}
\end{equation}
where the gradient moduli $\mathrm{B}_{\lambda}$ and
$\mathrm{C}_{\lambda}$ are found as
\begin{equation}
\frac{1}{2}\mathrm{B}_{\lambda} = \pi \rho^4 \frac{ (\nabla \mu_{\lambda})^2}{4} 
\left(\Sigma_{\text{hom}}^\parallel(\lambda, \mu_{\lambda}) + 5
\frac{\Gamma}{\rho}  \frac{\mu_{\lambda}}{\lambda} \right) \quad \text{and} \quad  \mathrm{C}_{\lambda}
= - \pi \rho^4 \frac{
	\Gamma}{2 \rho}  \frac{\mu_{\lambda}^2}{\lambda} \nabla \mu_{\lambda},
\label{eq:gradientmoduli}
\end{equation}
\end{subequations}
as announced in~(\ref{eq:RelaxedEnergy_announce})
and~(\ref{eq:gradientmoduli_announce}).  In the absence of surface
tension, $\Gamma = 0$, the modulus $\mathrm{C}_{\lambda}$ cancels and
the second term in $\mathrm{B}_{\lambda}$ vanishes: the prediction
of~\citet{Audoly-Hutchinson-Analysis-of-necking-based-2016} is
recovered, see equations 2.28a and 2.28b in their paper.  A
distinctive feature of our 1d strain gradient
model~(\ref{eq:RelaxedEnergy}), is that both the potential
$W_{\text{hom}}(\lambda)$ and the gradient moduli
$\mathrm{C}_{\lambda}$ and $\mathrm{B}_{\lambda}$ are nonlinear
functions of the strain $\lambda$ that capture both the material and
geometric nonlinearities present in the hyper-elastic cylinder model
which we started from.  By contrast, existing 1d models proposed in
the literature typically replace the functions
$W_{\text{hom}}(\lambda)$, $\mathrm{B}_{\lambda}$ and
$\mathrm{C}_{\lambda}$ by some expansions about a critical value of
$\lambda$ corresponding to the onset of localization.

Given any particular material model, one can classify the homogeneous
solutions $\mu_{\lambda}$ by solving the transverse equilibrium
equation~(\ref{eq:nlCylinderEquil_simple_announce}), and then derive
the 1d strain energy functional $\Phi_{\text{1d}}[\lambda]$ explicitly.  For
constitutive laws such that the catalog $\mu_{\lambda}$ is not
available in closed analytical form, $\mu_{\lambda}$ can be tabulated
numerically as a function $\lambda$ and $\Gamma$: this makes it
possible to evaluate numerically all the coefficients entering in the
non-linear equations of equilibrium~(\ref{eq:NLEq}).  For the
incompressible neo-Hookean model, considered
in~\citet{taffetani2015beading, taffetani2015elastocapillarity,
xuan2017plateau}, an explicit expression of the 1d strain energy
functional $\Phi_{\text{1d}}[\lambda]$ in~(\ref{eq:full1dModelAsDerived}) is
derived in appendix~\ref{a:incompressibleCase}.

To derive the equilibrium equations for the 1d model, one has to
impose that the total energy is stationary with respect to
perturbations in $\lambda$.  However, this variational problem is
ill-posed due to the presence of the boundary term
$[C_{\lambda}\,\lambda']_{0}^{L}$ appearing in $\Phi_{\text{1d}}$, as can be
checked.  This can be interpreted by the fact that the variational
structure of the problem is broken when higher-order terms are
discarded.  There are two possible ways around this difficulty.  The
first one is simply to ignore the boundary terms, \emph{i.e.}, to set
$C_{\lambda}=0$; this approximation is reasonably accurate (but not
exact, because we are also restricting the virtual perturbations
arbitrarily) for solutions such that $\lambda'$ cancels on both sides
of the domain, as the ones we consider in the sequel.  The second
approach is rigorous, but also slightly more complex, and involves
changing the definition of the centroid $z(S)$, so as to make the
boundary terms go away; this amounts to modify the expression of the
quantity $\mathrm{B}_{\lambda}$ in the 1d model, \emph{i.e.}, to use a
a quantity $\mathrm{D}_{\lambda}$ instead that differs by an
additional term, see equation~(\ref{eq:gradientmodulus_newBC}) in the
appendix.  The first approach is used in the numerical simulations
shown in the next section.  The second approach is documented in
appendix~\ref{a:boundaryTerm}, for future reference.  For all the
numerical simulations shown in the next section, we have verified that
the approximate and rigorous approaches yield curves that are very
similar: their respective curves can hardly be distinguished in any of
the plots.

We now proceed to derive the equations of equilibrium, assuming
$C_{\lambda}=0$.  The dimension reduction has been carried out so far
without considering any loading.  The equilibrium of the cylinder
subjected to a tensile force $F$ applied at its ends is governed by
the total potential energy of the system $\Phi_{\text{1d}}[\lambda] -F
\int_0^L \lambda(S) \mathrm{d}S$.  In the absence of any kinematic
constraint, the Euler-Lagrange equations characterizing the
equilibrium are found as
\begin{equation}
\begin{array}{c}
W_{\mathrm{hom}}'(\lambda) + \frac{1}{2} \frac{\mathrm{d} 
\mathrm{B}_{\lambda}}{\mathrm{d} \lambda}(\lambda(S))\, 
\lambda^{\prime \, 2} - \frac{\mathrm{d}}{\mathrm{d}S}\left(\mathrm{B}_{\lambda} \lambda'(S)\right) -F = 
0,\\
\\
\lambda'(S)=0 \quad \text{for} \quad S=0 \quad \text{and} \quad S=L.
\end{array}
\label{eq:NLEq}
\end{equation}
If, however, the position of the endpoints is prescribed, the external
force of the problem $F$ becomes an unknown; this unknown is set by
the condition that the average stretch $\lambda$ is consistent with
the end-to-end distance imposed by the boundary conditions.

\section{Numerical results and discussion}
\label{s:results}

In this section, we solve the 1d nonlinear boundary value
problem~(\ref{eq:NLEq}) numerically for different values of the
surface tension $\Gamma$, of the aspect-ratio $\epsilon$ and for
different types of boundary conditions.  The behavior of the system is
controlled by the dimensionless parameters $\epsilon$ and
$\overline{\Gamma} = \frac{\Gamma}{\rho G}$.  Localization is possible
when the unregularized energy $W_{\mathrm{hom}}$ is not convex, which
happens when $\overline{\Gamma} > \overline{\Gamma}_{\text{c}}$; the
critical dimensionless surface tension is
$\overline{\Gamma}_{\text{c}} = \sqrt{32}$.

We analyze two types of geometries in the forthcoming sections, and
compare the predictions of the 1d model to finite-element simulations
from earlier work~\citep{taffetani2015beading, xuan2017plateau}.  This
allows us to test the accuracy of our 1d gradient model and to
characterize its range of validity.  In~\S\ref{s:results}\ref{ssec:XuanCompare}, we derive a bifurcation
diagram when the axial force $F$ is varied, keeping the geometric and
physical parameters fixed.
The diagram is typical of a propagative instability, with a plateau
for the axial force $F$ associated with the propagation of a localized
interface sweeping the length of the cylinder.  Next,
in~\S\ref{s:results}\ref{ssec:taffetaniCompare}, we analyze the case
of fixed endpoints, such that the end-to-end distance is fixed to
$Z(L) = L$, and we vary the parameter $\overline{\Gamma}$.

In all the numerical simulations, we use an incompressible neo-Hookean
material model: we use the
expressions~(\ref{eq:1d-nHincompressible-Whom})
and~(\ref{eq:moduliBC}) from appendix~\ref{a:incompressibleCase} for
the terms appearing in the non-linear equilibrium
equation~(\ref{eq:NLEq}).

The 1d non-linear boundary-value problem~(\ref{eq:NLEq}) can be solved
numerically by
quadrature~\citep{Audoly-Hutchinson-Analysis-of-necking-based-2016},
but we prefer to use arc-length continuation instead.  To do so, we
use the AUTO-07p
library~\citep{Doedel-Champneys-EtAl-AUTO-07p:-continuation-and-bifurcation-2007}.
We solve~(\ref{eq:NLEq}) on a domain covering one half of the
cylinder, assuming a symmetry condition $\lambda'(L/2)=0$ at the
center;  
we focus attention on the first buckling mode in this half-domain,
which represents the second buckling mode of the full domain $(0,L)$
(the first mode in the full domain is anti-symmetric).  The other
buckling modes can be analyzed similarly.
Numerical simulations of the 1d model are considerably faster than
typical finite-element simulations in axisymmetric geometry: the
entire phase diagrams shown in figures~\ref{fig:BifurcGamma6}
and~\ref{fig:Taffetanicompare} can be generated in a few
seconds on a personal computer.

\subsection{Varying force, keeping elasto-capillary properties fixed}
\label{ssec:XuanCompare}

\begin{figure}
	\centering
	\includegraphics[width=13cm]{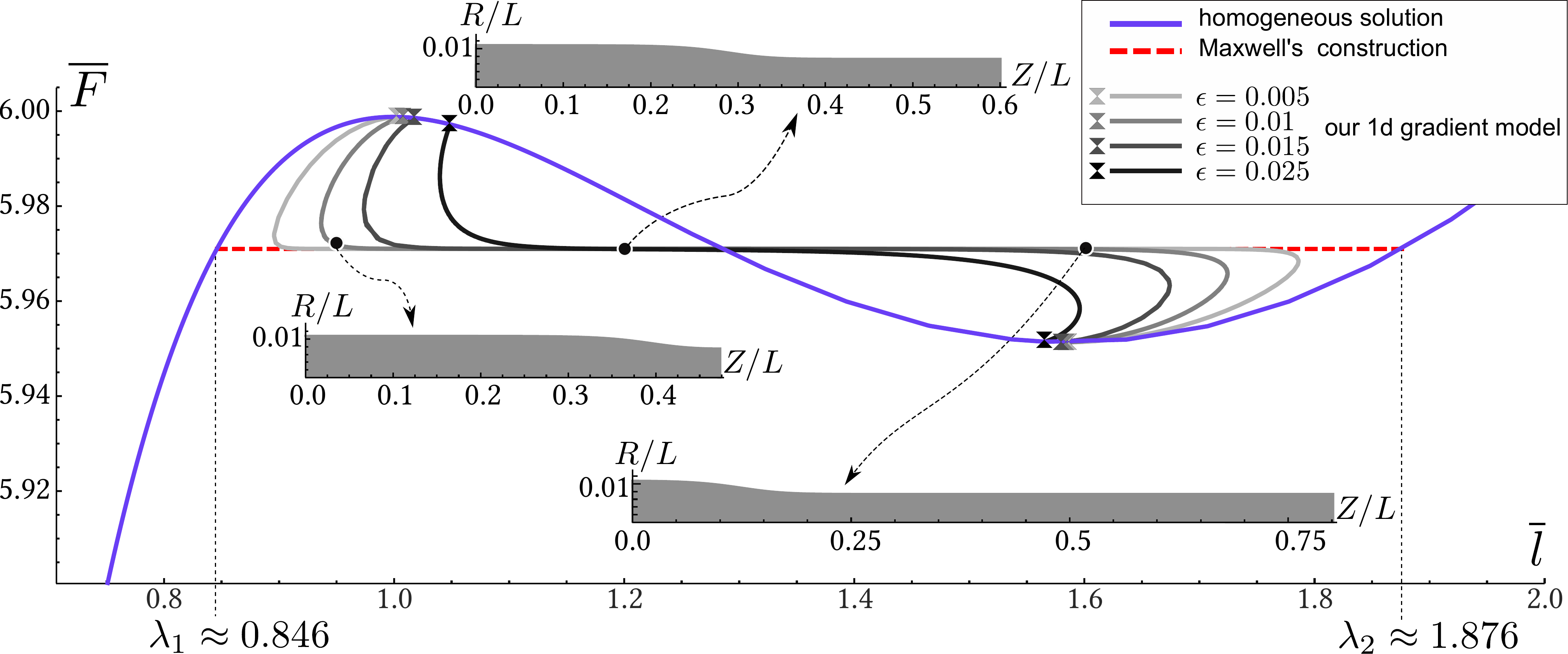}
	\caption{Bifurcation diagram predicted by the 1d gradient model in
	the $(\overline{l},\overline{F})$-plane for $\overline{\Gamma}=6$,
	where $\overline{l} = Z(L)/L$ is the average stretch ratio and
	$\overline{F} = F/( \pi G \rho^2)$ is the scaled force.  We
	consider symmetric modes with $\lambda'(S=L/2)=0$.  The gray
	curves were obtained by solving the 1d model using a continuation
	method for different aspect-ratios $\epsilon \in\{0.005, 0.01,
	0.015, 0.025\}$ (from light gray to black shades).  The
	bifurcation condition~(\ref{eq:bifurc}) is represented by the
	hourglass symbols.  The curve corresponding to homogeneous
	solutions, \emph{i.e.}, to a constant function $\lambda$, is drawn
	in blue; Maxwell's construction is represented by the dotted red
	line.  Insets: current configuration for $\epsilon=0.01$ at
	$\overline{l} = \{0.95, 1.2,1.6\}$ (only the intersection of the
	half-plane $\Theta=0$ and of the half-cylinder $S\leq L/2$ is
	shown: the $Z/L$-axis is the axis of symmetry while the
	endpoint on the right is the center of the symmetric neck).}
	\label{fig:BifurcGamma6}
\end{figure}

Localization is associated with a loss of convexity of the
non-regularized potential $W_{\text{hom}}(\lambda)$.  The
non-convexity manifests itself by the fact that the loading curve
$W'_{\mathrm{hom}}(\overline{l}=\lambda)=F$ corresponding homogeneous
solutions has an up-down-up shape in the plane $(\overline{l},F)$ when
$\overline{\Gamma}>\overline{\Gamma}_{\mathrm{c}}=\sqrt{32}=5.66$,
where $\overline{l}=Z(L)/L=\lambda$ is the (homogeneous) axial stretch
ratio: the curve for homogeneous solutions is plotted in blue in
figure~\ref{fig:BifurcGamma6} for $\overline{\Gamma}=6$.  The
continuous curves with the different shades of gray correspond to
solutions of the 1d strain gradient model, using finite values of the
aspect-ratio $\epsilon$.  These gray curves bifurcate from the blue
curve slightly after (respectively, before) the blue curve attains
Consid\`ere's point where the force is maximum (respectively,
minimum): this is a classical size effect captured by strain gradient
models. 
Slender cylinders tend to bifurcate closer to the Consid\`ere's point
of maximum force, while bifurcation is delayed for more stubby
cylinders. 

Using the 1d gradient model, the bifurcation load $\lambda$ can be
predicted by the implicit equation, see for
instance~\citet{Audoly-Hutchinson-Analysis-of-necking-based-2016}
\begin{equation}
\frac{\mathrm{d}^{2}W_{\text{hom}}}{\mathrm{d}\lambda^{2}}(\lambda) = 
-
\frac{4 \pi^2}{L^2} \mathrm{B}_\lambda.
\label{eq:bifurc}
\end{equation}
Numerical solutions of this equation for different values of
$\epsilon$ are represented by the hourglass symbols in
Figure~\ref{fig:BifurcGamma6}; they match accurately the bifurcation
points of the branches calculated by the continuation method.

When following the bifurcation curves starting from the bifurcation
point, the simulations capture a zone of localized axial stretch that
grows progressively in amplitude while the axial force decreases
(initiation); these solutions are unstable by standard symmetry
exchange arguments, until the fold point where the axial forces starts
to increase again.  Further down the curve, the axial force converges
to a Plateau value, independent on the value of $\epsilon$.  The
applied stretch can then be increased at a constant axial force, while
the interface sweeps the length of the cylinder (propagation).
Deformed shapes of the cylinder illustrating this evolution are drawn
in Figure~\ref{fig:BifurcGamma6} for $\epsilon = 0.01$.

The propagative behavior has been documented in other localized
instabilities such as elastic necking or the bulging of elastic
membranes~\citep{G83,Kyriakides-Chang-The-initiation-and-propagation-of-a-localized-1991}
and it can be interpreted as a phase transformation
process~\citep{chaterHutch, xuan2017plateau,
Lestringant-Audoly-A-diffuse-interface-model-2018}.  In this view,
equilibrium solutions made of phases of constant strain are connected
by sharp interfaces; the coexistence of these solutions is predicted
by the 1d energy at order $\epsilon^{0}$, see
equation~(\ref{eq:1Dmodel0}).  Consider two of such homogeneous phases
in equilibrium, with axial stretch $\lambda_1$ and $\lambda_2$, and
occupying fractions $f$ and $1-f$ of the initial domain (as measured
by their extent in the Lagrangian domain $0\leq S\leq L/2$),
respectively: the condition that the energy is stationary with respect
to $\lambda_{1}$, $\lambda_{2}$ and $f$ writes
\begin{align}
\begin{split}
&\frac{\partial W_{\text{hom}}(\lambda)}{\partial \lambda} - F = 0 \quad \text{for} \quad \lambda = \lambda_1 \quad \text{and} \quad \lambda = \lambda_2,\\
&W_{\text{hom}}(\lambda_1)-F\lambda_1 = W_{\text{hom}}(\lambda_2)-F\lambda_2.
\end{split}
\label{eq:Maxwell}
\end{align}
This system of equations can be solved for $\lambda_{1}$,
$\lambda_{2}$ and $F$, which yields Maxwell's plateau where two
phases can be in equilibrium in an infinite
domain~\citep{MaxwellNature, xuan2017plateau}.
The plateau is shown by the dotted red line in the diagram in
Figure~\ref{fig:BifurcGamma6}: it provides an accurate prediction
of the axial force observed during the propagation phase, when
interfaces are well localized.

\begin{figure}
	\centering
	\includegraphics{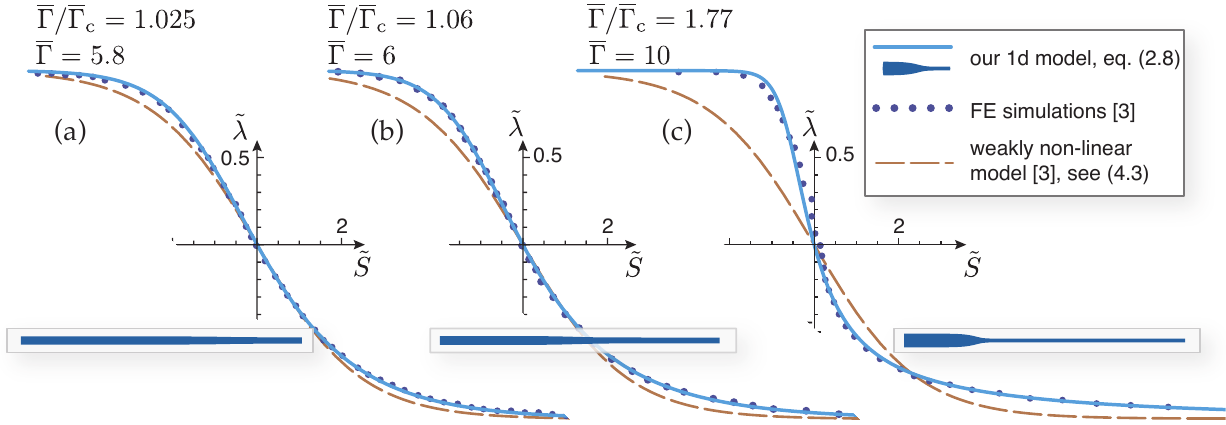}
	\caption{Solution for the interface in the propagation regime for
	different values of the rescaled surface tension
	$\overline{\Gamma} \in \{5.8, 6, 10\}$: solutions of the 1d
	gradient model with $\epsilon = 0.01$ and an average stretch ratio
	$\overline{l} = 1.2$ (light blue lines), finite element
	calculations in axisymmetric geometry from~\citet{xuan2017plateau} (dark blue dots), and weakly
	non-linear gradient model from Xuan \& Biggins (dashed brown
	curve).  The actual shape of the bar predicted by the 1d gradient
	model is shown in the overlays in dark blue.}
	\label{fig:Bigginscompare}
\end{figure}

The strain gradient model goes one step beyond Maxwell's construction
by allowing the spatial distribution of the two phases, the width and
the detailed shape of the interface to be determined.  In
Figure~\ref{fig:Bigginscompare}, we compare the deformed interface
predicted by the 1d gradient model (continuous blue line) with that
predicted by finite-element simulations of the full (3d-axisymmetric
hyper-elastic) model~(\ref{eq:NLElasticCylinderEnergy_simple})
available from~\citet{xuan2017plateau} (blue dots).  We plot the scaled
stretch ratio $\tilde{\lambda}(S) = \frac{2}{\lambda_2 -
\lambda_1}\left(\lambda(S) - \frac{\lambda_1 + \lambda_2}{2}\right)$,
where $\lambda_1$ and $\lambda_{2}$ are the transformation stretch
ratios as determined by Maxwell's construction~(\ref{eq:Maxwell}), as
a function of the scaled axial coordinate $\tilde{S} =
\frac{S-S_{\mathrm{ct}}}{\rho}\sqrt{\overline{\Gamma}-
\overline{\Gamma}_{\text{c}}}$; here $S_{\mathrm{ct}}$ is the
Lagrangian coordinate at the center of the neck.  In these scaled
variables, the prediction of Xuan and Biggins' weakly non-linear
analysis~\citep{xuan2017plateau} writes, see equation~(14) in their
paper,
\begin{equation}
	\tilde{\lambda}(\tilde S)= -\tanh
	\frac{2^{3/4}\tilde{S}}{\sqrt{17}}\qquad\qquad
	\textrm{(for 
	$|\overline{\Gamma}-\overline{\Gamma}_{\mathrm{c}}|\ll 1$),}
\end{equation}
and it is shown by the dashed brown line in
Figure~\ref{fig:Bigginscompare}.
This weakly non-linear analysis captures the shape of the interface
close to bifurcation, see Figure~\ref{fig:Bigginscompare}a.  However,
the agreement deteriorates in the post-bifurcation regime, the
prediction on the scaled stretch $\tilde{\lambda}$ being
underestimated by $\sim 10\%$ locally when $\overline{\Gamma}$ is 6\%
over the bifurcation threshold $\overline{\Gamma}_{\mathrm{c}}$, see
Figure~\ref{fig:Bigginscompare}(b).  By contrast, the 1d strain gradient
model derived here remains highly accurate over a significantly larger
range of values of $\overline{\Gamma}$, see
Figure~\ref{fig:Bigginscompare}(c) in particular.
	
\subsection{Varying elasto-capillary properties, keeping endpoints fixed}
\label{ssec:taffetaniCompare}

\begin{figure}
	\centering
	\includegraphics{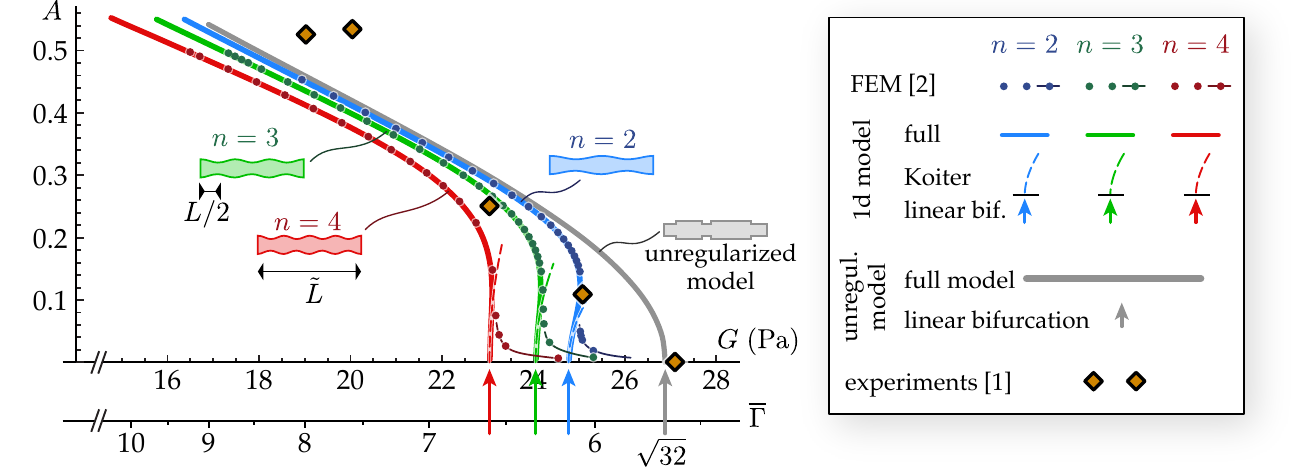}
	\caption{Necking in a cylinder having fixed endpoints: comparison
	of the predictions of the 1d model to the experiments of~\citet{mora2010capillarity} and to the finite-element
	simulations of~\citet{taffetani2015beading}.  In all
	simulations, an incompressible neo-Hookean model is used: based on
	the experimental parameters from~\citet{mora2010capillarity}, the
	shear modulus $G$ is connected to the scaled surface tension
	$\overline{\Gamma}$ by $G=152\;\mathrm{Pa}/\overline{\Gamma}$
	(double horizontal axis).  The amplitude $A$ of the modes
	comprising $n=2$, $3$, or $4$ full wavelengths are plotted (blue,
	green and red curves and data-points, respectively).  The
	amplitude $A$ is defined as the maximum
	minus the minimum radius, divided by twice the initial radius.
	The gray curve is the prediction of the unregularized model.  The
	agreement between the predictions of the 1d model (thick bright
	curves) and the finite-element simulations (dark dots) is
	excellent far from the bifurcation threshold; close to threshold,
	the finite-element simulations are affected by numerical
	imperfections (dark dots connected by thin dark curves).  The
	buckling threshold (arrows) and the weakly non-linear prediction
	obtained by the Koiter method (dashed parabolas) derived from the
	1d model agree well with its non-linear solutions; the instability
	is found to be sub-critical.}
	\label{fig:Taffetanicompare}
\end{figure}

In the experiments of~\citet{mora2010capillarity}, the amplitude
of necking in cylinders having fixed endpoints has been measured in
gels having different shear moduli $G$.  In these experiments, the
surface tension is
$\Gamma=36.5\,10^{{-3}}\,\mathrm{N}\,\mathrm{m}^{-1}$ and the initial
radius is $\rho=0.24\,\mathrm{mm}$, hence a scaled surface tension
$\overline{\Gamma} = \Gamma/(\rho\,G) = (152\;\mathrm{Pa})/G$.  With
the aim to interpret these experiments, Taffetani and Ciarletta
carried out non-linear finite element simulations of incompressible
neo-Hookean cylinders subjected to surface
tension~\citep{taffetani2015beading}, and investigated the post-buckled
equilibria of the mode with $n$ wavelengths, \emph{i.e.}, having
wavelength $\tilde{L}/n$, where $\tilde{L}=20\;\mathrm{mm}$ is the
actual length of the cylinder used in the experiments. A comparison with the predictions of our 1d model
is proposed in Figure~\ref{fig:Taffetanicompare}.  The 1d model was
simulated by the same method as earlier in
Figure~\ref{fig:BifurcGamma6}, except that the parameter
$\overline{\Gamma}$ was varied, and the force $\overline{F}$ was
treated as an unknown set by the condition
$\int_{0}^{L/2}\lambda(S)\,\mathrm{d}S = L/2$.  In addition, the 1d
model was solved numerically in a domain $0\leq S \leq L/2$
representing just a half wave of the buckling mode, \emph{i.e.}, a
fraction $1/(2n)$ of the length $\tilde{L}$ of the cylinder in the
experiments: $L=\tilde{L}/n$.  With this choice, the buckling modes
represented in the insets from Figures~\ref{fig:BifurcGamma6}
and~\ref{fig:Taffetanicompare} match.  Accordingly, we used in the 1d
model an aspect-ratio $\epsilon = \rho/L = n\,\rho/\tilde{L}=
0.012\,n$, where $\rho/\tilde{L}=0.012$ is the aspect-ratio in the
experiments.

Far from the bifurcation threshold, the agreement in
Figure~\ref{fig:Taffetanicompare} between the 1d model (bright thick
curves) and the finite-element simulations (dots) is excellent; this
confirms the accuracy of the 1d model in the deeply post-bifurcated
regime, as seen already in Figure~\ref{fig:Bigginscompare}.  Close to
the bifurcation threshold, the agreement is significantly poorer but
this is because the finite-element solution is affected by the
imperfections that were introduced numerically to trigger buckling, as
is evident from the `heels' near the horizontal axis (dark thin curves
connecting the dots); no such imperfections need to be introduced in
the continuation method used to solve the 1d model.  In fact, since
the 1d model has been derived based on the assumption that the strain
gradient $\lambda'(S)$ is small, it is even more accurate close to
threshold than far from threshold, even though the comparison to the
finite element solution is not meaningful close to threshold.

We have carried out a weakly non-linear analysis of the 1d model to
relate the dimensionless surface tension $\overline{\Gamma}$ to the
buckling amplitude $A$ of the bifurcation mode $\lambda(S) =
1-2\,A\,\cos(2\,\pi\,S/L)+\mathcal{O}(A^{2})=1-2\,A\,\cos(2\,n\,\pi\,S/\tilde{L})
+\mathcal{O}(A^{2})$ close to threshold (the coeffient $-2\,A$ here
warrants consistency with the amplitude $A$ defined in the legend in
terms of the local radius $\rho\,\mu(S) = \rho\,\lambda^{-1/2}(S)$
reconstructed by the incompressibility condition).
The weakly non-linear expansion is relatively straightforward to
derive, and very similar to that presented
in~\citet{Lestringant-Audoly-A-diffuse-interface-model-2018} for the
case of axisymmetric membranes; the details are not given here.  The
result is
\begin{subequations}
	\label{eq:KoiterExpansion-1dmodel}
\begin{equation}
	\overline{\Gamma}=\overline{\Gamma}_{\mathrm{c}}(n) +
\frac{\overline{\Gamma}''_{\mathrm{K}}(n)}{2}\,A^{2}+\mathcal{O}(A^{4}),
	\label{eq:weaklyNonlinearExpansion}
\end{equation}
where the critical parameter $\overline{\Gamma}_{\mathrm{c}}(n)$ is
such that
$W_{\mathrm{hom}}''(\lambda)=-\frac{4\,\pi^{2}}{\tilde{L}^{2}/n^{2}}\,B_{\lambda}$
at $\lambda=1$, as earlier in equation~(\ref{eq:bifurc}), and the
coefficient $\overline{\Gamma}''_{\mathrm{K}}(n)$ given by the Koiter
method writes
\begin{equation}
	\overline{\Gamma}''_{\mathrm{K}}(n) = 
	\frac{3 (4 (-8 + \overline{\Gamma})^2 - 
   4 \overline{k}_n^2 (24 - 83 \overline{\Gamma} + 13 \overline{\Gamma}^2) + 
   \overline{k}_n^4 (9 - 28 \overline{\Gamma} + 
      60 \overline{\Gamma}^2))}{16 \overline{k}_n^2 (-1 + 
	  \overline{k}_n^2) \overline{\Gamma}},
	  \quad
	  \overline{k}_{n} = 2\pi\,\frac{n\,\rho}{\tilde{L}}
\end{equation}
\end{subequations}
The
predictions of this weakly non-linear analysis are shown by the dashed
curves in the figure, and agree very well with the non-linear
simulations of the 1d model.  For the set of parameters used here, we
find $\overline{\Gamma}''_{\mathrm{K}}(n)<0$, implying that the
bifurcation is a sub-critical (discontinuous) pitchfork bifurcation:
along the bifurcated curves in figure~\ref{fig:Taffetanicompare}
(light thicker curves) and for an increasing amplitude $A$, the
parameter $\overline{\Gamma}$ initially decreases close to bifurcation
but the branch goes through an inflexion point soon after and
$\overline{\Gamma}$ increases with $A$.

A weakly non-linear analysis of an incompressible neo-Hookean cylinder
with surface tension in axisymmetric geometry was carried out
in~\citet{taffetani2015beading}.  For the same set of parameters as
those used in Figure~\ref{fig:Taffetanicompare}, the authors concluded
that the bifurcation is super-critical (continuous), which is at odds
with our own conclusion.  Our previous analysis of bulging in
axisymmetric
membranes~\citep{Lestringant-Audoly-A-diffuse-interface-model-2018}
shows that the 1d model captures amplitude equations such
as~(\ref{eq:weaklyNonlinearExpansion}) correctly at the dominant
order: the assumption $|\lambda'|\ll 1/L$ is well satisfied close to
threshold.  The weakly non-linear analyses based on the 1d model in
equation~(\ref{eq:KoiterExpansion-1dmodel}) and that reported
in~\citet{taffetani2015beading} should give exactly the same result.
We are confident that our result~(\ref{eq:KoiterExpansion-1dmodel}) is
correct as (\emph{i}) it accurately matches the non-linear solutions
close to threshold---such a verification could not be made
in~\citet{taffetani2015beading} due to the presence of the numerical
imperfections, see the `heels' in
figure~\ref{fig:Taffetanicompare}---, (\emph{ii}) the weakly
non-linear expansion based on the 1d model is considerably simpler
than that based on the full model, and therefore less prone to errors,
and (\emph{iii}) a close examination of the finite-element results
reveals a jump in the amplitude (a gap in the blue dots around
$G=25\;\mathrm{Pa}$ is visible in figure~\ref{fig:Taffetanicompare}),
which suggests that a discontinuity, not entirely suppressed by the
smoothing effect of the numerical imperfections, is indeed present.

Far from threshold, the bifurcation branches for $2\leq n\leq 4$ are
all well approximated by the curve shown in gray.  The gray curve has
been obtained from the unregularized model as follows: for any value
of $\overline{\Gamma}$, the transformation stretches $\lambda_{1}$ and
$\lambda_{2}$ have been calculated using Maxwell's construction, see
equation~(\ref{eq:Maxwell}), and the necking amplitude has been
defined based on the two phases in equilibrium as
$A=(\lambda_{2}^{{-1/2}}-\lambda_{1}^{-1/2})/2$.  In this
unregularized model, the interfaces are treated as discontinuities and
the detailed distribution of the coexisting phases along the length of
the cylinder remains undefined, but this is of no importance here.
The unregularized model gives similar predictions as the 1d gradient
model when the amplitude is larger than $A\stackrel{>}{\sim}0.3$: this
is because the buckling mode, which is initially evenly distributed
along the entire length of the cylinder, localizes quickly in the
post-bifurcation regime, thereby mimicking the predictions of the
unregularized model.  Overall, the unregularized model, despite being
quite simple, captures the general trend in the experimental
data-points fairly well.

In the bifurcation diagram in Figure~\ref{fig:Taffetanicompare}, the
Koiter expansion appears to have a very limited range of validity.
This is caused by the quick localization of the bifurcation mode in
the post-bifurcation regime, a phenomenon that is not captured by the
Koiter expansion (a similar issue arises in spherical shell
buckling~\citep{Audoly-Hutchinson-Localization-in-spherical-shell-2020}
in an even more severe form).  This fast localization phenomenon is
not captured by the weakly non-linear approaches that have been used
in earlier work on elasto-capillary necking.  By contrast, our 1d
model accurately captures localization, by retaining the relevant
nonlinearity in both the unregularized potential
$W_{\mathrm{hom}}(\lambda)$ and in the second gradient modulus
$B_{\lambda}$.

\section{Conclusion}
\label{s:conclusion}

We derived a 1d strain gradient model for the elasto-capillary necking
of soft hyper-elastic cylinders.  A generic, transversely symmetric
elastic constitutive law has been considered, which includes isotropic
materials as a particular case; the case of an incompressible
neo-Hookean material has been considered as an application.  The model
has been derived starting from a kinematic ansatz for the sake of
brevity, but the particular form of the displacement used as a
starting point can be justified from a systematic reduction
method~\citep{LESTRINGANT2020103730}.  It can also be checked that the
solution thus constructed satisfies the equations of equilibrium at
the dominant orders.  The resulting model is asymptotically correct in
the limit where the ratio $\epsilon$ of the radius $\rho$ to the
typical scale of variation of the longitudinal stretch $\lambda(S)$ is
small, $\epsilon\ll 1$.

Our model consists of a non-convex strain energy density depending on
the axial stretch ratio $\lambda(S)$ at leading order $\epsilon^{0}$,
regularized by a term depending quadratically on the strain gradient
$\lambda'(S)$ and non-linearly on $\lambda(S)$.  This regularizing
term is formally of order $\epsilon^{2}$.  Expressions for all the
coefficients entering the 1d strain energy functional have been
derived in terms of the cylinder's radius and of the hyper-elastic
constitutive behavior.  It is a distinctive and
crucial feature of our 1d model that the non-linear dependence of the
energy on $\lambda(S)$ is retained.  This warrants accurate
predictions in a broad range of parameters, as shown by the comparison
with finite element simulations of the full 3d finite-strain
elasticity problem.

Besides being significantly easier to solve numerically, the 1d model
is easily amenable to linear and weakly nonlinear bifurcation
analyses.  It also reveals the deep analogy with other propagative
instabilities and with phase transitions: the 1d model which we
derived is akin to the diffuse-interface model introduced by van der
Waals for the analysis of the liquid-vapor phase
transition~\citep{vdwaals94}.  Despite its simplicity, it is
remarkably accurate, even beyond the onset of localization where it is
mathematically justified.  This unexpected accuracy is manifest on the
third plot in Figure~\ref{fig:Bigginscompare}, when the bifurcation
parameter is as large as
$\overline{\Gamma}/\overline{\Gamma}_{\mathrm{c}} = 10/5.7 = 1.77$, as
well as in Figure~\ref{fig:Taffetanicompare}.  A similar, unexpectedly
broad domain of validity has been reported with the 1d strain gradient
model derived for the analysis of the bulging of axisymmetric
membranes~\citep{Lestringant-Audoly-A-diffuse-interface-model-2018}.
There is probably a common explanation for these nice surprises but we
could not identify it.  We hope that this intriguing fact will be
investigated further.

\vspace{1em}
\noindent\textbf{Ethics statement.} This work did not involve any ethical 
issue.\\[.2em]
\noindent\textbf{Data accessibility statement.} This work does not have any experimental data.
\\[.2em]
\noindent\textbf{Competing interests statement.} We have no competing interests.
\\[.2em]
\noindent\textbf{Authors' contributions.} Both authors have equally contributed to all aspects of the work.
\\[.2em]
\noindent\textbf{Funding.} There has been no dedicated funding for this work.
\\[.2em]
\noindent\textbf{Acknowledgments.}  We would like to thank Serge Mora for sharing pictures of his experiments.

\appendix

\section{Elimination of the boundary terms}
\label{a:boundaryTerm}

\renewcommand{\theequation}{\thesection.\arabic{equation}}

The dimension reduction produces boundary terms
$[C_{\lambda}\,\lambda']_{0}^{L}$ in the energy
functional~(\ref{eq:RelaxedEnergy}).  These boundary term make the
variational problem of equilibrium ill-posed.  In this appendix, we
fix this issue by introducing an alternate definition of the centroid
$z(S)$ of the cross-sections.  When expressed in terms of the new
centroid, the energy functional has no boundary terms, and the
equations of equilibrium can be obtained variationally.  This change
of unknown amounts to discard the boundary term
$\mathrm{C}_{\lambda}$, and to add another contribution to the
coefficient $\mathrm{B}_{\lambda}$.

Moving the boundary term in~(\ref{eq:RelaxedEnergy}) under the
integration sign yields
\[
     \Phi_{\text{1d}}[\lambda] 
	 = \int_0^L \left( W_{\text{hom}} (\lambda (S)) + \frac{1}{2}  \left(
     \mur{B}_{\lambda (S)} + 2 \frac{\mathd \mur{C}_{\lambda}}{\mathd \lambda}
     (\lambda (S)) \right) \lambda^{\prime 2} (S) + \mur{C}_{\lambda (S)}
     \lambda'' (S) \right) \mathd S +\mathcal{O} (L \varepsilon^3).
\]
With a view of eliminating the second derivative of $\lambda$ in this energy functional, we change the definition of the centroid to
\begin{equation}
z_{\star} (S) = z (S) + f (\lambda (S)) \lambda' (S),
\label{eq:newCentroid}
\end{equation}
where the function $f (\lambda)$ will be specified later.  With the
aim to motivate this change of variable, we observe that it is akin to
switching from the uniform averaging in~(\ref{eq:averageZ}), to a
{\tmem{weighted}} average with weight $g(R)$ in the cross-section,
\emph{i.e.}, $z_{\star} = \int_0^L g (T) Z (S, T) 2 \mathpi T \mathd
T$.  Indeed, the latter yields, in view of the curvature effect,
$z_{\star} \sim z+ \mathcal{A} (S) \int_0^L g (T) t (T) 2
\mathpi T \mathd T \sim z + \lambda' (S) \int_0^L g (T) t
(T) 2 \mathpi T \mathd T$, which is indeed of the same form as~(\ref{eq:newCentroid}).

With the new centroid definition~(\ref{eq:newCentroid}), the apparent stretch becomes
\[ \lambda_{\star} (S) = \frac{\mathd z_{\star}}{\mathd S} = \frac{\mathd
   z}{\mathd S} + \frac{\mathd f}{\mathd \lambda} \lambda^{\prime 2} + f
   \lambda'' = \lambda (S) + \frac{\mathd f}{\mathd \lambda} (\lambda (S))
   \lambda^{\prime 2} (S) + f (\lambda (S)) \lambda'' (S).\]
As both terms $\frac{\mathd f}{\mathd \lambda} \lambda^{\prime 2}$
and $f \lambda''$ are of order $\varepsilon^2$, this relation can be
inverted as
$
\lambda (S) = \lambda_{\star} (S) - \left( \frac{\mathd f}{\mathd \lambda}
   (\lambda_{\star} (S)) \lambda_{\star}^{\prime 2} (S) + f (\lambda_{\star}
   (S)) \lambda_{\star}'' (S) \right) +\mathcal{O} (\varepsilon^4)$.
In terms of $\lambda_{\star}$, the energy functional writes
\[ \begin{array}{lll}
     \Phi_{\text{1d}}[\lambda_\star] & = & \int_0^L \left[ W_{\text{hom}} \left(
     \lambda_{\star} (S) - \frac{\mathd f}{\mathd \lambda} (\lambda_{\star}
     (S)) \lambda_{\star}^{\prime 2} (S) - f (\lambda_{\star} (S))
     \lambda_{\star}'' (S) \right) \cdots \right.\\
     &  & \nobracket \nobracket \hspace{4em} \left. + \frac{1}{2}  \left(
     \mur{B}_{\lambda_{\star} (S)} + 2 \frac{\mathd \mur{C}_{\lambda}}{\mathd
     \lambda} (\lambda_{\star} (S)) \right) \lambda^{\prime 2} (S) +
     \mur{C}_{\lambda_{\star} (S)} \lambda_{\star}'' (S) \mathd S \right]
     +\mathcal{O} (L \varepsilon^3),\\
     & = & \int_0^L \left[ W_{\text{hom}} (\lambda_{\star} (S)) + \frac{1}{2}
     \left( \mur{B}_{\lambda_{\star} (S)} + 2 \frac{\mathd
     \mur{C}_{\lambda}}{\mathd \lambda} (\lambda_{\star} (S)) - 2 \frac{\mathd
     W_{\text{hom}}}{\mathd \lambda} (\lambda_{\star} (S))  \frac{\mathd
     f}{\mathd \lambda} (\lambda_{\star} (S)) \right) \lambda_{\star}^{\prime
     2} (S) \right. \cdots\\
     &  & \nobracket \nobracket \hspace{4em} \left. + \left(
     \mur{C}_{\lambda_{\star} (S)} - \frac{\mathd W_{\text{hom}}}{\mathd
     \lambda} (\lambda_{\star} (S)) f (\lambda_{\star} (S)) \right)
     \lambda_{\star}'' (S) \right] \mathd S +\mathcal{O} (L \varepsilon^3).
   \end{array} \]
With the particular definition
\begin{equation}
f (\lambda) = \frac{\mur{C}_{\lambda}}{\frac{\mathd W_{\text{hom}}}{\mathd
		\lambda} (\lambda)},
\label{eq:functionf}
\end{equation}
the term proportional to $\lambda_{\star}'' (S)$ cancels in the new
energy functional.  This corresponds to a re-definition of the
centroid~(\ref{eq:newCentroid}) and of the stretch measure as
\begin{equation}
	 z_{\star} = z + \frac{\mur{C}_{\lambda}}{\frac{\mathd
   W_{\text{hom}}}{\mathd \lambda}} \lambda',\qquad
   \lambda_{\star} = \lambda + \frac{\mathd}{\mathd \lambda} \left(
   \frac{\mur{C}_{\lambda}}{\frac{\mathd W_{\text{hom}}}{\mathd \lambda}}
   \right) \lambda^{\prime 2} + \frac{\mur{C}_{\lambda}}{\frac{\mathd
   W_{\text{hom}}}{\mathd \lambda}} \lambda''
   \textrm{.}
\end{equation}
Note that both $\lambda$ and $\lambda_{\star}$ are of order 1, while 
the correction is small, of order $\epsilon^{2}$.

Inserting the expression of $f$ in the expression of the energy above, we find
\begin{subequations}
\begin{gather}
\Phi_{\text{1d}}[\lambda_\star] = \int_0^L \left( W_{\text{hom}} (\lambda_{\star} (S)) +
\frac{1}{2}  \mur{D}_{\lambda_{\star} (S)} \lambda_{\star}^{\prime 2} (S)
\right) \mathd S +\mathcal{O} (L\, \varepsilon^3),
\label{eq:gradientmodel_newBC}
\\
\intertext{where}
\mur{D}_{\lambda_{\star}} = \mur{B}_{\lambda_{\star}} + 2
\mur{C}_{\lambda_{\star}}  \frac{\frac{\mathd^2 W_{\text{hom}}}{\mathd
		\lambda^2} (\lambda_{\star})}{\frac{\mathd W_{\text{hom}}}{\mathd \lambda}
	(\lambda_{\star})}.
\label{eq:gradientmodulus_newBC}
\end{gather}
\end{subequations}

A comparison of the original energy~(\ref{eq:RelaxedEnergy}) and the
one just derived shows that the boundary terms in the energy can be
discarded provided the elastic modulus $B_{\lambda}$ is replaced with
$D_{\lambda}$.

\section{1d model for an incompressible neo-Hookean material}
\label{a:incompressibleCase}

The strain energy of a quasi-incompressible neo-Hookean material with a shear modulus $G$, subjected to
an equi-biaxial strain of axial stretch $\lambda$ and radial strain 
$\mu$, can be written as
\[ w (\lambda, \mu) = \frac{G}{2} \left( \lambda^2 + 2 \, \mu^2 \right) +
   \frac{1}{2 \, \eta} \, \left( \lambda \, \mu^2 - 1 \right)^2,
\]
where $\eta \ll 1$ is a penalization parameter enforcing
inextensibility.  There are different ways to this energy that are
equivalent in the limit $\eta\to 0$, and we picked a simple one.
The transverse equilibrium~(\ref{eq:nlCylinderEquil_simple_announce}) 
yields
\begin{equation}
  2 \, G \, \mu + \frac{1}{\eta} \, \left( \lambda \,
  \mu^2 - 1 \right) \, 2 \, \mu \, \lambda + \frac{2 \,
  \Gamma}{\rho} \, \lambda = 0. \label{eq:1d-incompr-pressure}
\end{equation}
From~(\ref{eq:sigmahom}), the axial Piola-Kirchhoff stress is
$ \Sigma^{\parallel} = \frac{1}{\lambda} \, \frac{\partial w}{\partial
   \lambda} = G + \frac{1}{\eta} \, \left( \lambda \, \mu^2 - 1
   \right) \, \frac{\mu^2}{\lambda}$.
Inserting the hydrostatic pressure $\frac{1}{\eta}
\, \left( \lambda \, \mu^2 - 1 \right)$ found
from~(\ref{eq:1d-incompr-pressure}), we have $\Sigma^{\parallel}
= \frac{1}{\lambda} \, \frac{\partial w}{\partial \lambda} = G \,
\left( 1 - \frac{\mu}{\lambda} \, \overline{\Gamma} - \,
\frac{\mu^2}{\lambda^2} \right)$.  In the incompressible limit, $\eta
\rightarrow 0$ and $\mu_{\lambda} = \lambda^{- 1 / 2}$, so that
\[ \Sigma^{\parallel}_{\text{hom}} (\lambda) = G \, \left( 1 -
   \frac{\overline{\Gamma}}{\lambda^{3 / 2}} - \, \frac{1}{\lambda^3}
   \right).\]

   The tangent shear modulus $K_{\mathrm{hom}}^{{STST}}(\lambda)$ can
   be found by considering a shear perturbation $\hat{\mathbf{E}} =
   \hat{E}_{ST}\,(\mathbf{d}_{S}\otimes \mathbf{d}_{T} +
   \mathbf{d}_{T}\otimes \mathbf{d}_{S})$ to the equi-biaxial strain
   $\mathbf{E}_{\mathrm{hom}}(\lambda,\lambda^{{-1/2}})$, as
   in~(\ref{eq:w-expansion-to-second-order-A}), by expanding the
   strain energy, and by identifying the result with
   with~(\ref{eq:w-expansion-to-second-order-B}). This yields
   \begin{equation}
   	K_{\text{hom}}^{STST}(\lambda)
	= G\,
\left(\lambda^{-2}+\overline{\Gamma}\,\lambda^{{-1/2}}\right).
   	\label{eq:KSTST-incompr}
   \end{equation}

The potential of the unregularized model reads from~(\ref{eq:Whom}) 
\begin{equation}
	W_{\text{hom}} (\lambda) = \frac{\mathpi \, \rho^2 \, G}{2}
   \, \left( \lambda^2 + \frac{2}{\lambda} + 4 \, \overline{\Gamma}
   \, \lambda^{1 / 2} \right),
	\label{eq:1d-nHincompressible-Whom}
\end{equation}
as announced earlier in equation~(\ref{eq:order0_biggins}), and in equation 2 in~\citet{xuan2017plateau}. 

The gradient of transverse stretch defined in~(\ref{eq:nablaNotation})
reads $\nabla \mu_{\lambda} = \frac{\mathd \mu_{\lambda}}{\mathd
\lambda} = \frac{\mathd \lambda^{- 1 / 2}}{\mathd \lambda} = -
\frac{1}{2 \, \lambda^{3 / 2}}$, and we find the gradient moduli in
the energy~(\ref{eq:RelaxedEnergy}) from~(\ref{eq:gradientmoduli}) as
\begin{equation}
\frac{1}{2} \, B_{\lambda} = \frac{\mathpi \, \rho^4 \,
	G}{16 \, \lambda^6} \, \left( \lambda^3 + 4 \,
\overline{\Gamma} \, \lambda^{3 / 2} - 1 \right), \qquad \qquad
C_{\lambda} = \frac{\mathpi \, \rho^4 \, G}{4 \, \lambda^{7
		/ 2}} \, \overline{\Gamma}.
\label{eq:moduliBC}
\end{equation}
The elimination of boundary terms using the method in
appendix~\ref{a:boundaryTerm} is carried out by
inserting~(\ref{eq:1d-nHincompressible-Whom}) and~(\ref{eq:moduliBC})
into~(\ref{eq:gradientmodulus_newBC}), which yields
\begin{equation}
  \frac{1}{2} \, D_{\lambda} = \frac{\mathpi \, \rho^4 \,
   G}{16 \, \lambda^6} \, \frac{(\lambda^3 - 1)^2 + 3 \,
   \overline{\Gamma} \, \lambda^{3 / 2} \, \left( 3 \,
   \lambda^3 + 1 \right) + 2 \, \overline{\Gamma}^2 \,
   \lambda^3}{\lambda^3 + \overline{\Gamma} \, \lambda^{3 / 2} - 1}.
   \label{eq:1d-nHincompressible-Dlambda}
\end{equation}
The relative change in second-gradient modulus $\left| \frac{D_{\lambda} -
B_{\lambda}}{B_{\lambda}} \right|$ is typically less that 10\% for
$\overline{\Gamma} \leq 10$ and $1 \leq \lambda \leq 2$.

\vspace{20px}

\bibliographystyle{plainnat}
\bibliography{CylinderWithSurfaceTension-papers}

\end{document}